\documentclass[12pt]{article}
\usepackage[table,xcdraw]{xcolor}
\usepackage{amsmath}
\usepackage{amsmath,bm}
\usepackage{url}
\usepackage{tikz}
\usetikzlibrary{shapes.geometric}
\usetikzlibrary{arrows}
\usepackage{amsmath}
\usepackage{amssymb}
\usepackage{hyperref}
\usepackage{rotating}
\usepackage{physics}
\usepackage{epsfig}
\usepackage{graphicx}
\usepackage{color}
\usepackage{cite}
\usepackage{subfig}
\usepackage{slashed}
\usepackage[font=small,labelfont=bf]{caption}

\newcommand{\beq}{\begin{equation}}
\newcommand{\eeq}{\end{equation}}
\newcommand{\ga}{\lower.7ex\hbox{$\;\stackrel{\textstyle>}{\sim}\;$}}
\newcommand{\la}{\lower.7ex\hbox{$\;\stackrel{\textstyle<}{\sim}\;$}}

\hypersetup{
    colorlinks = true,
    citecolor = {blue},
    linkcolor = {blue},
    urlcolor = {blue},
}

\begin{document}\sloppy 

\def\jcap{\ref@jnl{J. Cosmology Astropart. Phys.}}

\begin{flushright}
{\tt KCL-PH-TH/2020-38}, {\tt CERN-TH-2020-123}  \\
{\tt ACT-05-20, MI-TH-2020} \\
{\tt UMN-TH-3924/20, FTPI-MINN-20/27} \\
\end{flushright}


\begin{center}
{\bf {\large Non-Oscillatory No-Scale Inflation}}

\end{center}


\begin{center}{
{\bf John~Ellis}$^{a}$,
{\bf Dimitri~V.~Nanopoulos}$^{b}$,
{\bf Keith~A.~Olive}$^{c}$ and
{\bf Sarunas~Verner}$^{c}$}
\end{center}

\begin{center}
{\em $^a$Theoretical Particle Physics and Cosmology Group, Department of
  Physics, King's~College~London, London WC2R 2LS, United Kingdom;\\
Theoretical Physics Department, CERN, CH-1211 Geneva 23,
  Switzerland;\\
  National Institute of Chemical Physics \& Biophysics, R{\" a}vala 10, 10143 Tallinn, Estonia}\\[0.3cm]
{\em $^b$George P. and Cynthia W. Mitchell Institute for Fundamental
 Physics and Astronomy, Texas A\&M University, College Station, TX
 77843, USA};\\
{\em Astroparticle Physics Group, Houston Advanced Research Center (HARC),
 \\ Mitchell Campus, Woodlands, TX 77381, USA;\\ 
Academy of Athens, Division of Natural Sciences,
Athens 10679, Greece}\\[0.3cm]
{\em $^c$William I. Fine Theoretical Physics Institute, School of
 Physics and Astronomy, University of Minnesota, Minneapolis, MN 55455,
 USA}
 
 \end{center}

\vspace{0.3cm}
\centerline{\bf ABSTRACT}
\vspace{0.1cm}

{\small We propose a non-oscillatory no-scale supergravity model of inflation (NO-NO inflation) in which the inflaton does not oscillate at the end of the inflationary era. Instead, the Universe is then dominated by the inflaton kinetic energy density (kination). During the transition from inflation to kination, the Universe preheats instantly through a coupling to Higgs-like fields. These rapidly annihilate and scatter into ultra-relativistic matter particles, which subsequently dominate the energy density, and reheating occurs at a temperature far above that of Big Bang Nucleosynthesis. 
After the electroweak transition, the inflaton enters a tracking phase as in some models of quintessential inflation.
The model predictions for
cosmic microwave background observables are consistent with Planck~2018 data, and
the density of gravitational waves is below the upper bound from Big Bang Nucleosynthesis. We also find that the density of supersymmetric cold dark matter produced by 
gravitino decay is consistent with Planck~2018 data over the expected range of supersymmetric
particle masses.
 } 

\vspace{0.2in}

\begin{flushleft}
{August} 2020
\end{flushleft}
\medskip
\noindent

\newpage

\section{Introduction}
\label{sec:intro}

The theory of inflation is the most successful mechanism for explaining how the Universe became so homogeneous and isotropic on large scales~\cite{reviews}. The usual scenarios assume that, after the epoch of cosmic inflation, the scalar inflaton field rolled down to a minimum and started oscillating, ultimately decaying into elementary particles and reheating the Universe~\cite{reheating, nos}.

However, one may also consider an alternative mechanism of reheating, caused by non-perturbative effects arising from a  parametric resonance, known as preheating~\cite{preheating,stb, kt,Greene:1997fu} (see~\cite{prerev} for reviews). If the inflaton field is coupled to another scalar field, $h$, the large effective mass of the $h$ field changes non-adiabatically when the inflaton crosses the origin, causing explosive production of $h$ particles. 

The original models of preheating~\cite{preheating,stb, kt,Greene:1997fu} discussed the exponential particle production that occurs non-perturbatively due to inflaton field oscillations about an effective minimum. However, it was later shown by Felder et al.~\cite{instpreh} that preheating can occur instantaneously without  inflaton oscillations or a parametric resonance, if the coupling between the inflaton and the additional scalar field is sufficiently large. This discovery stimulated interest in non-oscillatory (NO) models of inflation~\cite{NOinfl}. 
These have suffered from the paucity of efficient reheating mechanisms other than gravitational particle production induced by the changing space-time metric~\cite{gravprod, Hashiba:2018iff, Haro:2018zdb}. One alternative possible reheating mechanism for NO models of inflation was to introduce a curvaton field~\cite{curvaton1, curvaton2}, which decays into thermalized radiation and reheats the Universe. Models with curvaton reheating do not need an interaction term between the inflaton field and another scalar field. 

In this paper, we focus our attention on NO models of inflation with instant preheating. 
Because the process of instant preheating is extremely efficient, only a small fraction of the inflaton energy is converted to the production of the heavy $h$ particles, which in turn rapidly annihilate and scatter into ultra-relativistic particles. After the period of instant preheating, the Universe is dominated by the kinetic energy density of the inflaton, which scales as $\propto a^{-6}$~\cite{kination}, until the energy density of ultra-relativistic products, which scales as $\propto a^{-4}$, starts dominating the Universe, resulting in reheating. 
A viable scenario of instant preheating requires that the Universe enters the radiation-dominated era significantly before the epoch of Big Bang Nucleosynthesis (BBN), which necessitates a reheating temperature $T_{\rm RH} \gg 1 \, \rm{MeV}$. On the other hand, in supersymmetric models too large a value of $T_{\rm RH}$ would bring the risk of excessive production of gravitinos whose decays would overproduce dark matter. 

The non-oscillatory model of instant preheating postulated in~\cite{instpreh} assumed that, after the inflationary epoch, when the inflaton field crosses the zero point and becomes negative, the effective scalar potential vanishes and the energy density of the inflaton consists only of the kinetic energy contribution. It was later shown that the mechanism of instant preheating works very well in models incorporating quintessential inflation, as originally introduced by Peebles and Vilenkin~\cite{PV},
which assume that the inflaton field slowly rolls toward zero potential after inflation,
and could explain the present-day dark energy.
Quintessential inflation with instant preheating has been studied in several different contexts, including 
the production of gravitational waves~\cite{giovannini,tash},
$\alpha$-attractor models of inflation~\cite{alpha2}, the primordial production of black holes~\cite{prepbh}, Gauss-Bonnet models~\cite{pregb}, and UV freeze-in models~\cite{preUV}. In these models the coupling between the inflaton and another scalar field is generally put in by hand, and the coupling strength is adjusted to avoid the overproduction of radiation created by instant preheating, which would otherwise cause the reheating temperature $T_{\rm RH}$ to be too large.

In this paper we consider a new scenario for instant preheating, derived from no-scale supergravity~\cite{no-scale, Ellis:1983sf, LN}.
In this scenario,
immediately after instant preheating the effective mass of the non-perturbatively produced particle $h$, which we treat as a proxy for the Higgs boson of the Standard Model (SM), is similar to the inflationary scale. These superheavy particles then annihilate and scatter into ultra-relativistic products. The Universe enters a period dominated by the inflaton energy density (kination) until the ultra-relativistic particles start to dominate the energy density of the Universe and reheating occurs.
The motion of the inflaton is slowed down during this radiation-dominated phase.
Then, after the electroweak phase transition, 
when the radiation density drops to the level of the inflaton potential, the universe enters a tracking phase, as in quintessence models~\cite{Steinhardt:1999nw}. 
The residual inflaton potential energy later acts as dark energy in the present-day Universe. 

We are motivated to consider models based on $\mathcal{N} = 1$ supergravity, because
it provides a natural framework that incorporates a viable dark matter candidate and also connects the inflaton to the SM fields at high scale. The theory of supergravity is characterized by the geometric properties of a non-trivial K\"ahler manifold, which can incorporate naturally the required effective interaction terms between the inflaton and the proxy Higgs field $h$. Further, we are motivated to study scenarios based on no-scale supergravity~\cite{no-scale, Ellis:1983sf, LN} because it avoids undesirable anti-de Sitter states and emerges in the effective low-energy limit of string compactification~\cite{Witten}. For some previous models combining the SM with no-scale models of inflation, see~\cite{ENO8, EGNO4, ENOV4, king2, egnno1, EGNNO23, EGNNO45, dgmo}.

We demonstrate in this paper the construction of a non-oscillatory, no-scale supergravity model of inflation (NO-NO inflation), whose predictions agree with the most recent cosmic microwave background (CMB) measurements~\cite{planck18, rlimit}. This model has a simple plateau inflationary potential, whose predictions are similar to those of the original Starobinsky model of inflation based on $R + R^2$ gravity~\cite{Staro} (see~\cite{ENO6, Avatars, ENOV1, ENOV2,KLno-scale,FKR,FeKR,adfs,king,eno9,others} for other Starobinsky-like models in the context of no-scale supergravity), and the general framework can be extended easily to models of no-scale attractors~\cite{ENOV3}, which we leave for future work. 
We note that plateau models of inflation with a minimum have also been studied in~\cite{tmodel, multitmodel}, where they were referred to as T-models.

The structure of this paper is as follows. In Section~II we describe our NO-NO
inflation scenario, first reviewing briefly relevant aspects of no-scale supergravity, then discussing
the effective potential of the model and equations of motion before analyzing
its predictions for CMB observables, which are compatible 
with the Planck~2018 data~\cite{planck18,rlimit}. 
Then post-inflationary preheating is studied in Section~III. An estimate of the reheating temperature is given in Section~IV, where the tracking regime and its conclusion is also discussed.
The abundance of gravitational waves is calculated in Section~V and found to be compatible with the
upper limit from Big Bang Nucleosynthesis, and the abundance of gravitinos and the density of supersymmetric dark matter produced by their decays are considered in
Section~VI.
Finally, Section~VII summarizes our results and discusses our conclusions.

\section{NO-NO Inflation}
\label{sec:non-osc}

\subsection{No-Scale Framework}
In this Section we illustrate how to construct non-oscillatory (NO) models of inflation in the non-minimal $\mathcal{N} = 1$ no-scale supergravity framework.~\footnote{For discussions of other $\mathcal{N} = 1$ no-scale supergravity models, see~\cite{no-scale, Ellis:1983sf, LN}.} In addition to the inflaton field, $\phi$, we include a scalar field, $h$, corresponding to the Higgs boson. The two scalars are coupled only through interactions induced by the supergravity Lagrangian. During inflation, 
the inflaton field has a large field value of $\sim \mathcal{O} (M_P)$, and $h \simeq 0$.~\footnote{We are using the reduced Planck mass, $M_P = 1/\sqrt{8 \pi G_N} \simeq 2.4 \times 10^{18}$ GeV. With the exception of providing units for numerical quantities, we will work in units where $M_P = 1$, and thus factor of $M_P$ will be omitted in most analytical expressions.}
When the inflaton field $\phi$ crosses the region near the null point, $\phi \simeq 0$, the field velocity, $\dot{\phi}$, is large and 
the mass of the Higgs boson changes nonadiabatically due to the effective interaction term between the $\phi$ and $h$ fields, and heavy particles are produced explosively through the preheating process, as we discuss in detail in the next Section.

We introduce the following K\"ahler potential
\begin{equation}
\label{kah1}
    K \; = \; -\alpha \log[1 - x \bar{x} - y \bar{y}] - \beta \log[1-z \bar{z}] \, ,
\end{equation}
where $x$ is associated with the inflaton field, $y$ with the Higgs field, and $z$ is an auxiliary field. The K\"ahler potential~(\ref{kah1}) parametrizes an $\frac{SU(2, 1)}{SU(2) \times U(1)} \times \frac{SU(1, 1)}{U(1)}$ coset manifold, and the curvature parameters $\alpha, \beta$ describe the characteristic geometry of the manifold.~\footnote{One could also consider a K\"ahler potential of the form
 $   K \; = \; -\alpha \log[1 - x \bar{x} - y \bar{y} - z \bar{z}]$,
 and construct similar non-oscillatory
scalar potentials, albeit with a different superpotential. The relation between these forms and the more standard form  $   K \; = \; -\alpha \log[T+\bar{T} - \phi_i \bar{\phi}_i]$ is explained in Appendix A.} We recall that the K\"ahler curvature is given by $R_{i \bar{j}} \equiv \partial_i \partial_{\bar{j}} \ln K_{i \bar{j}}$, where the K\"ahler metric is defined as $K_{i \bar{j}} \equiv \partial^2 K/\partial \Phi^i \partial \bar{\Phi}^{\bar{j}}$ and where the $\Phi_i$ are the complex chiral fields, and the Ricci scalar curvature can be expressed as
\begin{equation}
\label{curv}
    R = R_{i \bar{j}} K^{i \bar{j}} \, ,
\end{equation}
where $K^{i \bar{j}}$ is the inverse K\"ahler metric.

The scalar curvature for the K\"ahler potential~(\ref{kah1}) is $R = 4/\alpha + 2/\beta$, where we use the convention that $R > 0$ for a hyperbolic manifold and $R < 0$ for a spherical manifold. In our further discussion, we assume that $\alpha, \beta > 0$ and hence we have a hyperbolic manifold with $R > 0$.

The scalar $\mathcal{N} = 1$ supergravity action for the chiral fields is given by
\begin{equation}
\label{action}
    S=\int d^{4} x \sqrt{-g}\left(\mathcal{L}_{\rm{kin}}-V\right) \, ,
\end{equation}
where the scalar kinetic term takes the form
\begin{equation}
\label{kin}
\mathcal{L}_{\mathrm{kin}}=K_{i \bar{j}} \partial_{\mu} \Phi^{i} \partial^{\mu} \bar{\Phi}^{\bar{j}} \, ,
\end{equation}
and the effective scalar potential is given by
\begin{equation}
\label{potSG}
V=e^{G}\left[\frac{\partial G}{\partial \Phi^{i}} K^{i \bar{j}} \frac{\partial G}{\partial \bar{\Phi}^{\bar{j}}}-3\right] \, 
\end{equation}
where, in order to include interactions, we have introduced an extended K\"ahler potential including a superpotential $W$:
\begin{equation}
    G \equiv K + \ln W + \ln \overline{W} \, .
\end{equation}
We assume in our discussion of inflationary dynamics in the supergravity framework
a spatially-flat Universe, which is described by the Friedmann-Robertson-Walker (FRW) metric
\begin{equation}
    ds^2 = dt^2 - a(t)^2 d {\bf{x}}^2 \, .
\end{equation}
The classical equations of motion for the scalar fields are given by
\begin{equation}
\label{mot1}
    H^2 = \dot{N}^2 =\frac{1}{3} \left[K_{i \bar{j}} \dot{\Phi}^i \dot{\bar{\Phi}}^{\bar{j}} + V(\bf{\Phi}) \right] \, ,
\end{equation}
\begin{equation}
\label{mot2}
    \dot{H} = - K_{i \bar{j}} \dot{\Phi}^i \dot{\bar{\Phi}}^{\bar{j}} \, ,
\end{equation}
\begin{equation}
\label{mot3}
    \ddot{\Phi}^{i} + 3H \dot{\Phi}^i + \Gamma^{i}_{jk} \dot{\Phi}^j \dot{\Phi}^k + K^{i \bar{j}} \frac{\partial V}{\partial \bar{\Phi}^{\bar{j}}} = 0 \, ,
\end{equation}
where ${\bf{\Phi}} \equiv (x, y, z)$, $H = \dot{a}/a$ is the Hubble parameter, $\Gamma^{i}_{jk} = K^{i \bar{l}} \partial_{j} K_{k \bar{l}}$ are the Christoffel symbols, and $N$ is the number of e-folds. 

\subsection{Non-Oscillatory Effective Potential}
We now construct a non-oscillatory model of inflation starting with K\"ahler potential~(\ref{kah1}). We introduce a superpotential of the form
\begin{equation}
\label{sup1}
    W = \frac{m z}{\sqrt{3}} \left(1 - x^2 \right)^{b/2} \left(1 + x \right)^{c/2} + \mu y^2 \, ,
\end{equation}
where $m$ is inflationary scale and the $\mu$ term is associated with the electroweak 
symmetry-breaking scale. Although we consider, for simplicity, a single field associated with the Higgs field, the general framework can easily be extended to more realistic supersymmetric scenarios with two Higgs doublets as in the minimal supersymmetric extension of the SM.

In order to stabilize the fields in the tachyonic directions, we introduce the following higher-order correction term in the K\"ahler potential~(\ref{kah1}):~\cite{EKN3,Avatars,EGNO4}
\begin{equation}
    \label{stabz}
    K \supset - \beta \log \left[1 - z \bar{z} + \frac{|z|^4}{\Lambda_z^2} \right] \, ,
\end{equation}
where $\Lambda_z$ is a mass scale that is smaller than the Planck scale $M_P$. This higher-order correction term, which we view as a perturbation of the original no-scale structure,  does not spoil the inflationary potential and ensures that other fields remain fixed during inflation.~\footnote{For a related discussion of the flatness and stability conditions in supergravity, see~\cite{GomezReino:2006wv, Covi:2008cn}.} The higher-order term in the K\"ahler potential~(\ref{stabz}) stabilizes the field $z$ in the real and imaginary directions with $z = \bar{z} = 0$. Due to the symmetric structure of the superpotential~(\ref{sup1}), we find numerically that when $z = \bar{z} = 0$, the curvature in the imaginary field directions ${\rm{Im}} \, x$ and ${\rm{Im}} \, y$ is positive, with a minimum located at $ \langle {\rm{Im}} \, x \rangle = \langle {\rm{Im}} \, y \rangle = 0$, and no additional stabilization terms in the K\"ahler potential~(\ref{kah1}) are necessary. This ensures that the potential is minimized along $y = \bar{y}$ and $x = \bar{x}$.


Combining the superpotential~(\ref{sup1}) with the K\"ahler potential~(\ref{kah1}), we find the following effective scalar potential
\begin{equation}
    \label{pot1g}
    V  =  \frac{ \frac{1}{3} m^2(1 - x^2)^b (x + 1)^{c}}{\beta(1- y^2 - x^2)^\alpha} 
+ \frac{ \mu^2 y^2 (4 - 4 x^2 + y^2 (\alpha - 8+ (\alpha-2)^2 x^2) + y^4 (\alpha-2)^2)}{\alpha(1- y^2 - x^2 )^\alpha }
    \, .
\end{equation}
If we choose the following illustrative parameter values: $c = 4$, $b = \alpha = 2$, and $\beta = 1$, we obtain a simpler expression
\begin{equation}
    \label{pot1}
    V = \frac{ \frac{1}{3} m^2(x - 1)^2 (x + 1)^{6} + \mu^2 y^2 (2 - 3 y^2 - 2 x^2)}{(y^2 + x^2 - 1)^2} \, ,
\end{equation}
for which the scalar curvature $R = 4$.\footnote{If we used a K\"ahler potential parametrized by a $\frac{SU(3, 1)}{SU(3) \times U(1)}$ coset manifold, the choices $c = 4$, $\alpha = 2$, and $b = \alpha - 1$ would yield the same scalar potential, with $m/\sqrt{3} \to m/\sqrt{3/2}$.  } 

Because we expect $\mu \ll m$,
inflation is driven by the first terms in
Eq.~(\ref{pot1g}) and the numerator of Eq.~(\ref{pot1}).
The curvature parameter $\beta$ is associated with the auxiliary field $z$, so its value does not affect the inflationary dynamics nor, in turn, the cosmological observables $n_s$ and $r$.
It does, however, scale the inflaton potential, so that the quantity $m^2/\beta$ is fixed by the amplitude of the density fluctuations as discussed below. It should be noted that one can treat the parameters $(\alpha, \beta, b, c)$ in the effective scalar potential~(\ref{pot1g}) as free variables and build alternative non-oscillatory models similar to no-scale attractors, whose different values of these parameters lead to different values of the scalar tilt $n_s$ and tensor-to-scalar ratio $r$. However, not all parameter values are expected to accommodate sufficient preheating, because different values of the parameters change the steepness of the effective inflationary potential, which in turn affects the inflaton velocity that is directly related to the violation of adiabaticity. Here we focus on one particular relatively simple example as an existence proof, and we leave such considerations for future work. 

At the end of inflation, the inflaton field value $x$ keeps decreasing towards negative values and its energy density exhibits the typical kination scaling $\rho_{x} \propto a^{-6}$. 
After reheating, the Universe is dominated by radiation and the evolution of the inflaton slows, until the Universe enters a tracking period, where the radiation density, and the kinetic and potential energies of the inflaton evolve together. Eventually, the field $x$ approaches a field value of $-1$.\footnote{This asymptotic value follows from the canonical field parametrization~(\ref{can1}).} At later times, 
cold dark matter dominates the expansion, 
until very late times when the Universe becomes
dominated once again by the residual vacuum energy, with the slowly-evolving inflaton providing dark energy.

During inflation the Higgs field acquires
a large mass (as mentioned above and discussed further below), and we can safely assume that $y \simeq 0$ and neglect the kinetic mixing  between the fields in the Lagrangian~(\ref{kin}). This assumption is valid during the inflation, kination and preheating eras, but fails to hold for a brief period when $x \simeq y \simeq 0$. We find from numerical calculations that neglecting the kinetic mixing between the two fields does not affect the preheating mechanism.

\begin{figure*}[ht!]
    \centering
    \includegraphics[width=\textwidth]{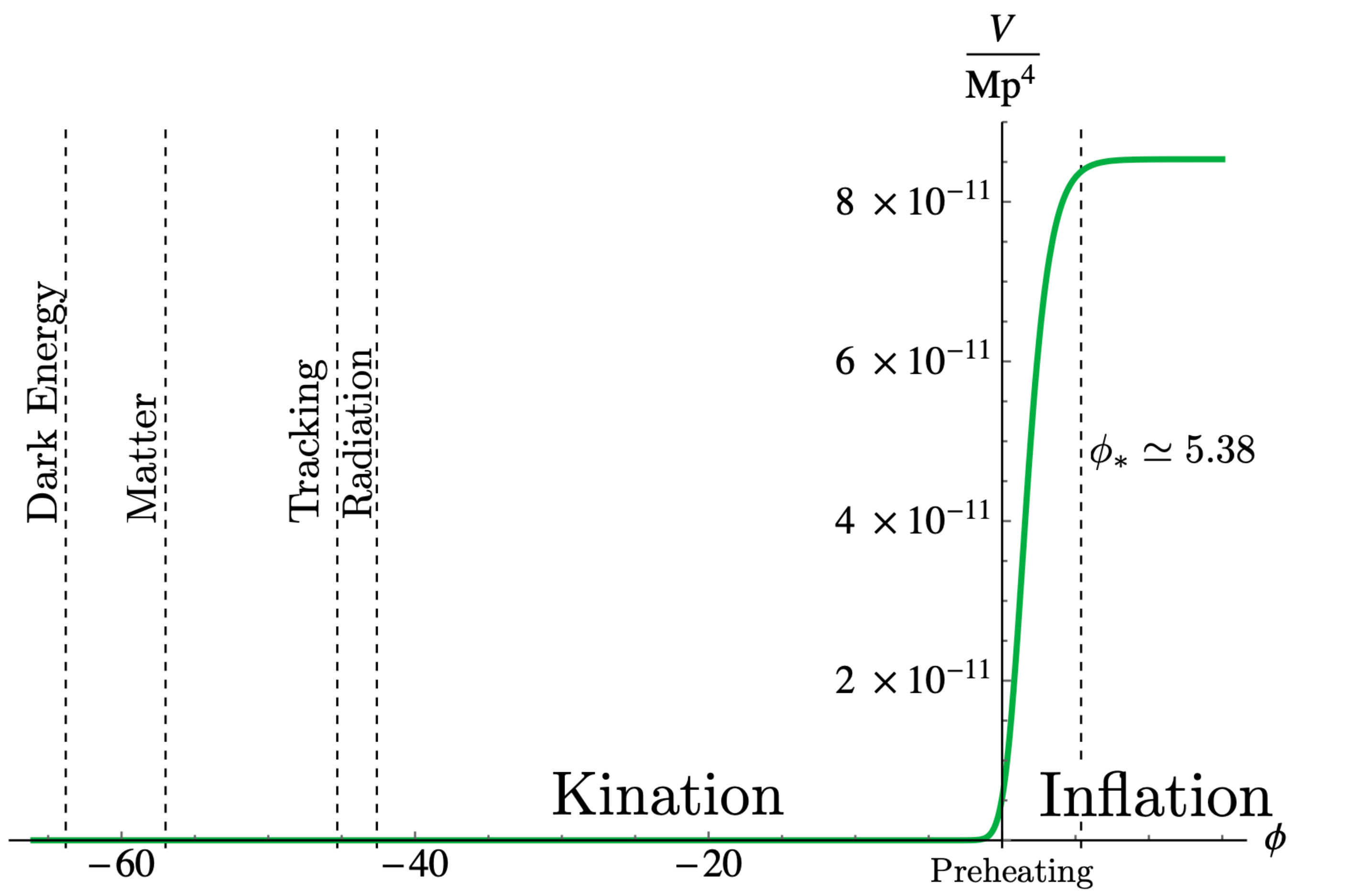}
    \caption{\it The scalar potential, $V_{\rm inf}(\phi)$, in the NO-NO model of inflation given by Eq.~(\ref{pot4}) with $m = 4 \times 10^{-6} \, M_P$. }
    \label{fig:potvsphi}
\end{figure*}

If we neglect the kinetic mixing terms in the Lagrangian~(\ref{kin}), the canonical parametrization for the fields $x$ and $y$ is given by
\begin{equation}
\label{can1}
    x = \tanh \left(\frac{\phi}{2} \right) \, ,
\end{equation}
\begin{equation}
\label{can2}
    y = \frac{h}{\sqrt{2 \left(\cosh{\phi} + 1 \right)}} \, ,
\end{equation}
where the canonically-normalized Higgs field, $h$, is coupled to the inflaton field, $\phi$. If we use the field redefinitions~(\ref{can1}, \ref{can2}), we can rewrite the scalar potential~(\ref{pot1}) as
\begin{equation}
\label{pot3}
    V \simeq V_{\rm{inf}} + \frac{1}{2} \left(\mu^2 + V_{\rm{inf}} \right)h^2 +\cdots \, ,
\end{equation}
where
\begin{equation}
\label{pot4}
    V_{\rm{inf}} = \frac{m^2}{3} \left(1 + \tanh {\frac{\phi}{2}} \right)^{4} \, ,
\end{equation}
and we have omitted Planck-suppressed higher-order terms other than those contributing to 
the effective Higgs mass:
\begin{equation}
\label{higgseff}
    m_{\rm h, eff}^2 = \frac{m^2}{3} \left(\tanh\left(\frac{\phi}{2} \right)  + 1 \right)^4 + \mu^2 \, .
\end{equation}
For $\phi \simeq 0$, we find $m_{h,{\rm eff}} \simeq m/\sqrt{3}$, since $m \gg \mu$.
The potential $V_{\rm inf}$ given by Eq.~(\ref{pot4}) with $m = 4 \times 10^{-6} \, M_P$ is shown
in Fig.~\ref{fig:potvsphi}.

In this approximation, the potential 
does not account for the VEV of $h$, 
as it does not include any effects due to 
supersymmetry breaking. In principle, the additional supersymmetry breaking sector may spoil the scalar potential at both early (inflationary) times and at late times. Indeed, even adding a simple constant to the superpotential greatly distorts the scalar potential in Eq.~(\ref{pot4}).  In order to avoid these potential problems, we can follow the treatment presented in~\cite{ENOV2, ENOV3, ENOV4} and include an adjustable parameter for supersymmetry breaking and dark energy without invoking any additional fields. A full treatment of supersymmetry breaking and the electroweak transition lies beyond the scope of this paper and will be considered in future work. 
However, a specific example in the basis (\ref{kah1}) is given in Appendix B.
Here we assume that the full theory contains supersymmetry breaking and generates a soft mass term for the Higgs field
that runs to a negative value at the electroweak scale~\cite{Ellis:1983sf}. 
When this and $D$-terms are included in the potential,
a VEV for $h$ is found.

\subsection{Inflation}
We use the slow-roll approximation in order to analyze the primary cosmological observables, namely the scalar tilt, $n_s$, and the tensor-to-scalar ratio, $r$. The slow-roll parameters
\begin{equation}
\label{slowroll1}
    \epsilon \equiv \frac{1}{2} \left(\frac{V'}{V} \right)^2 \, , \qquad \eta \equiv \frac{V''}{V}
\end{equation}
for the inflationary potential~(\ref{pot3}) are given by:
\begin{equation}
\label{slowroll2}
    \epsilon = \frac{8}{(1 + e^{\phi})^2} \, , \qquad \eta = \frac{4(4 - e^{\phi})}{(1 + e^{\phi})^2} \, .
\end{equation}
The total cosmological expansion is characterized by the number of e-folds:
\begin{equation}
    N_{*} \simeq - \int_{\phi_*}^{\phi_{\rm end}} \frac{1}{\sqrt{2 \epsilon}} d \phi \, ,
\end{equation}
where $\phi_*$ denotes the inflaton field value at the Hubble horizon crossing. Using~(\ref{slowroll2}), we obtain
\begin{equation}
    N_{*} \simeq \frac{(e^{\phi_*} + \phi_*)-(e^{\phi_{\rm end}} + \phi_{\rm end})}{4} \, .
\end{equation}
We define the end of inflation to occur when $\epsilon = 1$,~\footnote{A more precise definition for the end of inflation would be $\epsilon_H = 1$, where $\epsilon_H = - \dot{H}/{H^2}$, and in this case we find $\phi_{\rm{end}} \simeq 0.14$, corresponding to an insignificant difference in $\phi_*$.} and (\ref{slowroll2}) yields $\phi_{\rm end} \simeq 0.6$. For the nominal choices $N_* = 50, 55, 60$, we find that $\phi_* \simeq 5.28, 5.38, 5.47$. 

In order to connect the NO-NO model of inflation to the most recent CMB 2018 data~\cite{planck18,rlimit}, we use the following relations for the scalar tilt and scalar-to-tensor ratio in terms of the slow-roll parameters:
\begin{equation}
    n_s \simeq 1 - 6 \epsilon + 2 \eta, \qquad r \simeq 16 \epsilon \, .
\end{equation}
For our inflationary potential~(\ref{pot4}) and the choices $N_* = 50, 55, 60$, we find $n_s \simeq 0.9594, \, 0.9631, \, 0.9660$ and $r \simeq 0.0032, 0.0027, 0.0023$, which are consistent with the most recent CMB data~\cite{planck18,rlimit}.

We determine the inflationary scale $m$ from the amplitude of the scalar density fluctuations~\cite{planck18}, given by 
\begin{equation}
    A_s = \frac{V}{24 \pi^2 \epsilon},
\end{equation}
where $A_s \simeq 2.1 \times 10^{-9}$. For the nominal values of $N_* = 50, \, 55, \, 60$, we find $m \simeq 4.4 \times 10^{-6} \, M_P, \, 4.0 \times 10^{-6} \, M_P, \, 3.7 \times 10^{-6} \, M_P$. The key results for our model of NO-NO inflation are summarized in Table~\ref{tab1} below. In our further analysis we choose the representative value $N_* = 55$, and the corresponding values $\phi_* \simeq 5.38 \, M_P$ and $m \simeq 4 \times 10^{-6} \, M_P$.

\begin{table}[ht!]
\captionsetup{width=.65\textwidth}
\centering
\begin{tabular}{|c|c|c|c|}
\hline
$N_{*}$    & $50$                        & $55$                        & $60$                        \\ \hline \hline
$\phi_{*}$ & $5.28$                      & $5.38$                      & $5.47$                      \\ \hline
$m$        & $4.4 \times 10^{-6} \, M_P$ & $4.0 \times 10^{-6} \, M_P$ & $3.7 \times 10^{-6} \, M_P$ \\ \hline \hline
$n_s$      & $0.9594$                   & $0.9631$                   & $0.9660$                   \\ \hline
$r$        & $0.0032$                    & $0.0027$                    & $0.0023$                    \\ \hline
\end{tabular}
\caption{\it Parameters of the NO-NO inflationary potential~(\ref{pot4}) and CMB predictions for the specific choices $N_{*} = 50, 55, 60$.}
\label{tab1}
\end{table}

Since the effective mass of the Higgs field is similar to the inflationary scale during inflation, $m_{\rm{h, eff}} \simeq m$ as can be readily seen from (\ref{pot3}), the scalar potential in the $h$ direction is very steep, which leads to $h \simeq 0$. Therefore, in our model we do not have large Higgs field fluctuations, nor production of isocurvature perturbations \cite{Graziani:1988bp,NOinfl}.

Next, we use the classical equations of motion~(\ref{mot1} - \ref{mot3}) for the scalar fields to find numerical results for the time evolution of the model, which we use when discussing preheating in the next Section. In our numerical solutions and figures, time is measured in units $M_P^{-1}$, unless mentioned explicitly.

\begin{figure}[ht!]
    \centering
    \includegraphics[width=0.8\textwidth]{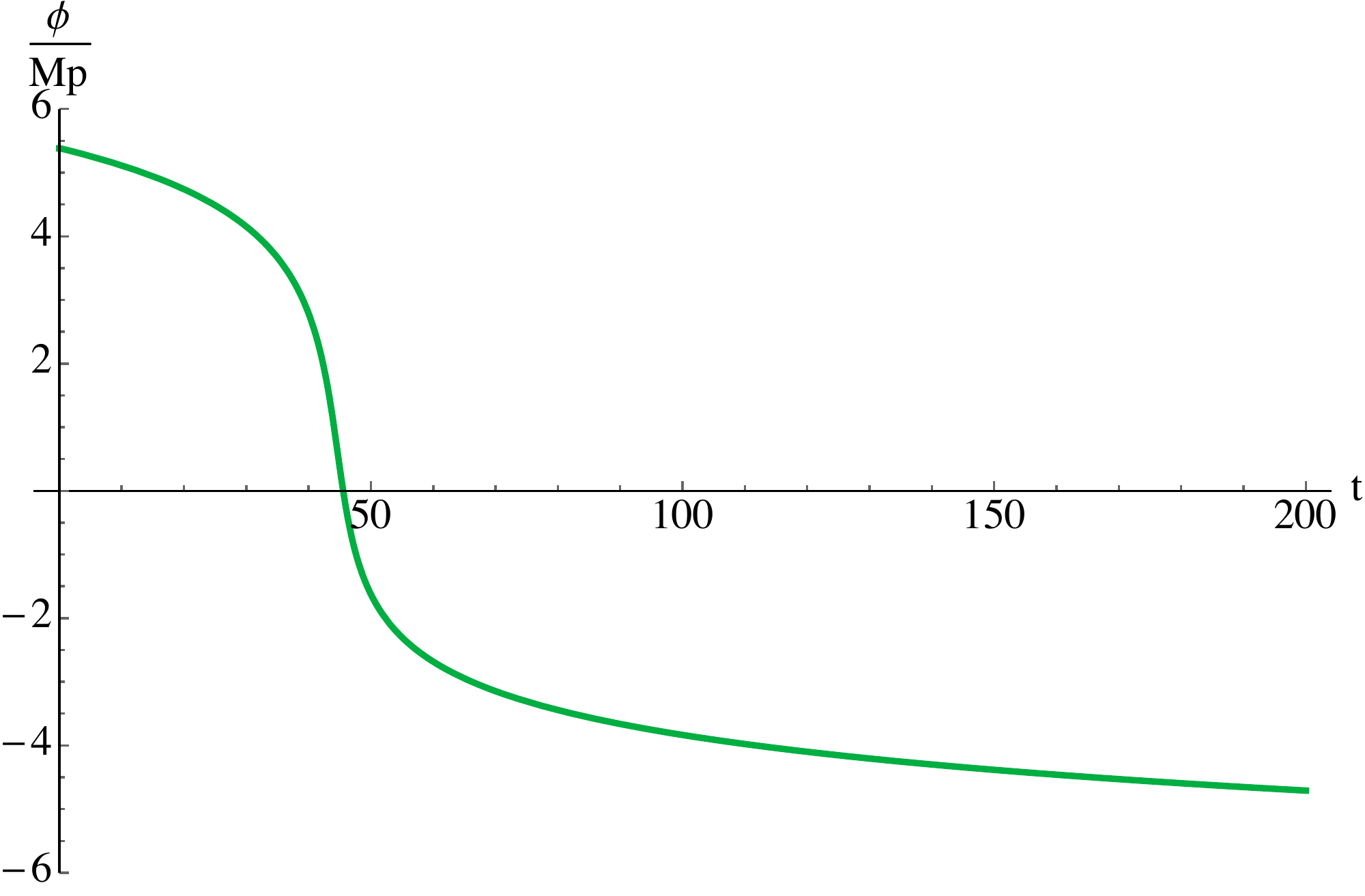}
    \caption{\it Evolution of the NO-NO inflaton field $\phi$ as a function of time in the units of $m^{-1}$.}
    \label{fig:plot1}
\end{figure}

\begin{figure}[ht!]
    \centering
    \includegraphics[width=0.8\textwidth]{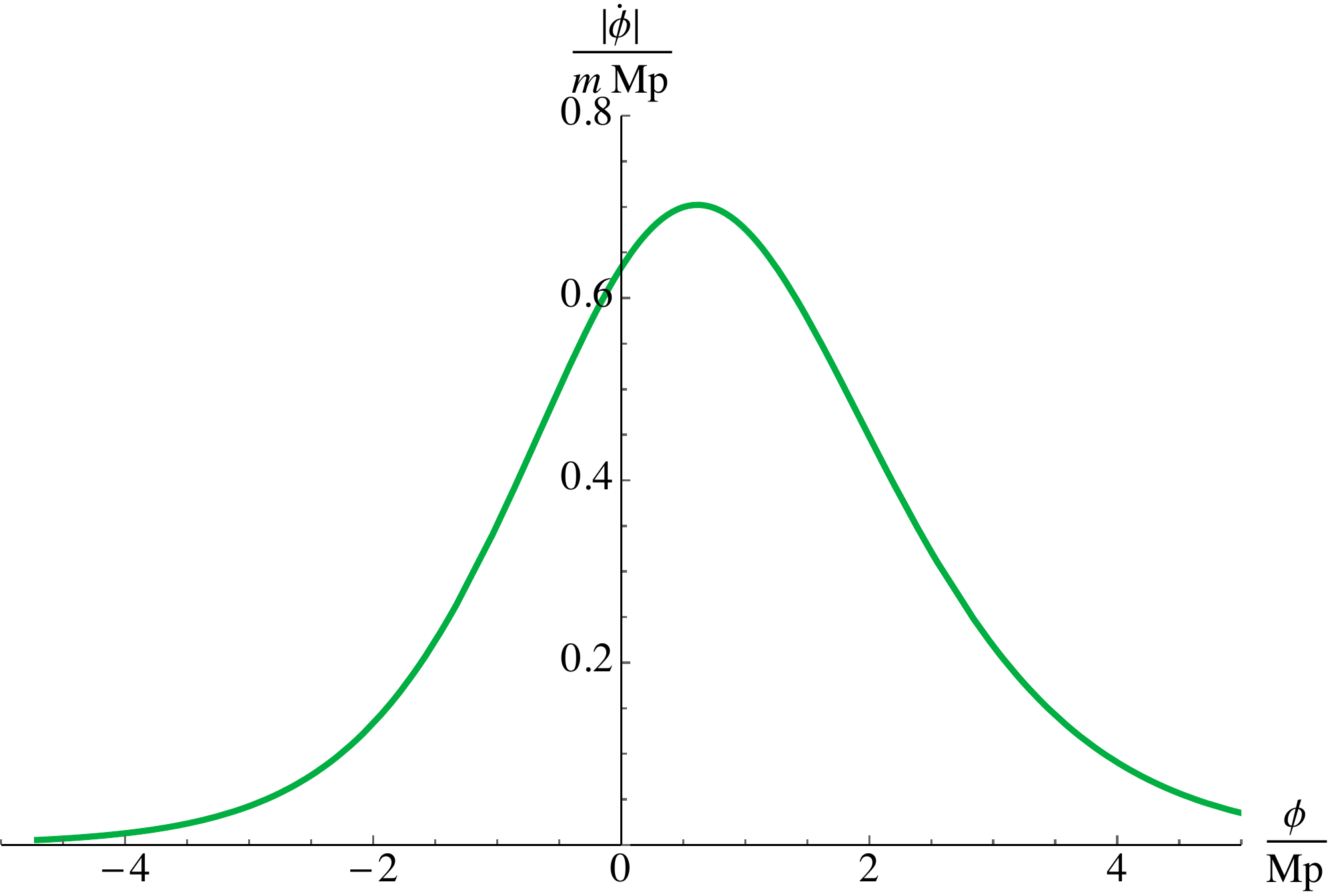}
    \caption{\it Evolution of the NO-NO inflaton velocity $|\dot{\phi}|$ as a function of inflaton field value.}
    \label{fig:plot2}
\end{figure}

We first show in Fig.~\ref{fig:plot1} the evolution of the inflaton field $\phi$ as a function of time.
The inflationary period ends when $\epsilon = 1$, or $\phi_{\rm{end}} \simeq 0.6 \, M_P$, and instant preheating begins. We then show in Fig.~\ref{fig:plot2} the evolution of the inflaton velocity $\dot{\phi}$, plotted as a function of the field value $\phi$. Our numerical calculations show that the inflaton velocity peaks right at the end of inflation, when $\phi \simeq 0.6 \, M_P$ and $|\dot{\phi}| \simeq 0.7 m \, M_P$, and that the kinetic energy of the inflaton becomes equal to its potential energy when $\phi \simeq -0.5 \, M_P$, when kination begins.


\begin{figure*}[ht!]
    \includegraphics[width=1.05\textwidth]{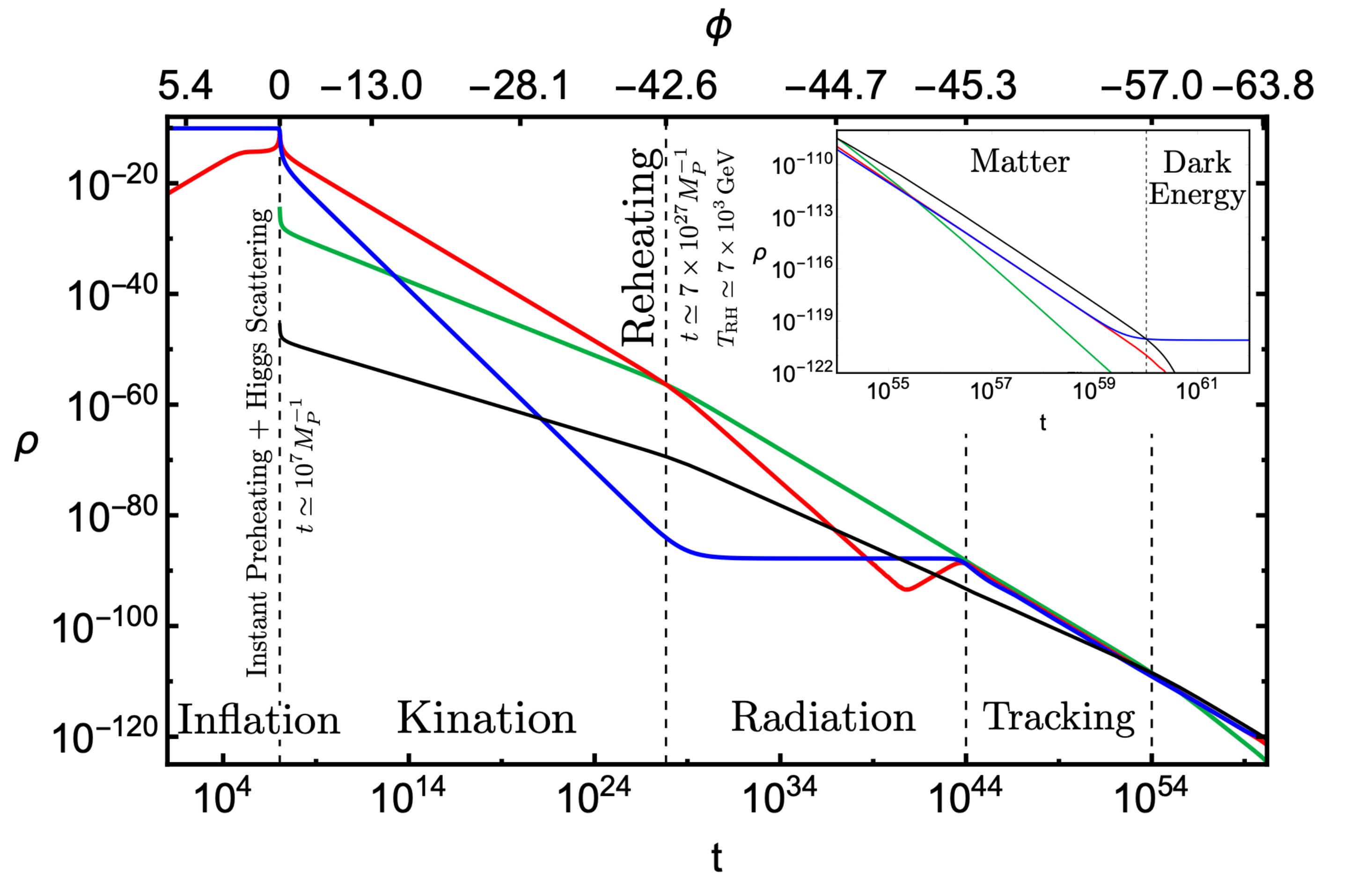}
    \caption{\it Evolution of the NO-NO inflaton and radiation energy densities as functions of time. The energy stored in the potential is depicted by the blue curve, the red curve shows the contribution of the inflaton kinetic energy,  the green and black curves corresponds to the radiation  and matter energy densities respectively.  Time is shown on the lower horizontal axis in units of $M_P^{-1}$, and some specific corresponding inflaton field values are given on the upper horizontal axis, in Planck units. The energy densities, $\rho$, are given in units of $M_P^4$. The Figure shows six distinct phases of evolution.  At early times, $t\le 10^7 \, M_P^{-1}$, the Universe is inflating, and the potential energy is relatively constant with a value given by $m^2 M_P^2/3$. As $\phi$ approaches $0$, its kinetic energy increases and dominates the energy density
    as the Universe enters a period of kination. When $\phi \simeq 0$, instant preheating occurs, the radiation
    bath is produced and begins to dominate at $t \simeq 7 \times 10^{27} \, M_P^{-1}$,
    which is defined as the moment of reheating. In the radiation-dominated phase, the inflaton slows to a near halt and the potential energy
    density becomes constant again until the radiation density,
    which continues to fall like $\rho_r \sim a^{-4} \sim t^{-2}$, becomes close to the inflaton potential and kinetic energy densities. At this time, $t \simeq  10^{44} \, M_P^{-1}$, the Universe enters a tracking phase that lasts until matter domination at $t \simeq 10^{54} \, M_P^{-1}$.  At the slightly later time of $t \simeq 10^{60} \, M_P^{-1}$, the vacuum energy which has been tuned to its current value $\sim 10^{-120} \, M_P^4$, begins to dominate as dark energy. 
    }
    \label{fig:hist}
\end{figure*}

After the inflaton crosses this point, the inflaton energy density becomes dominated by its kinetic energy contribution. This is shown as the red curve in Fig.~\ref{fig:hist}, which traces the history of the relevant energy densities as functions of time and the inflaton field value. When $\phi \simeq 0$, $t \simeq 10^7 \, M_P^{-1}$ (corresponding to $t \simeq 45 \, m^{-1}$ in Fig.~\ref{fig:plot1}) and the Universe becomes dominated by the inflaton kinetic energy.  The inflaton potential energy density (blue curve) then remains a subdominant contribution until
the Universe enters a tracking stage (discussed further below) at $t \simeq  10^{44} \, M_P^{-1}$. 
This result can be understood intuitively from the expression~(\ref{pot3}) for the effective scalar potential where, as $\tanh{\frac{\phi}{2}}$ approaches $-1$, a small value in the parentheses is raised to the fourth power, leading to a very small value of $V$. Therefore, when the inflaton passes the point $\phi \simeq -0.5 \, M_P$, we can use an approximation for the classical equations of motion $\ddot{\phi} + 3H \dot{\phi} \simeq 0$ and, as expected, the energy density of the inflaton field scales $\propto a^{-6}$. 

\begin{figure*}[ht!]
    \centering
    \includegraphics[width=\textwidth]{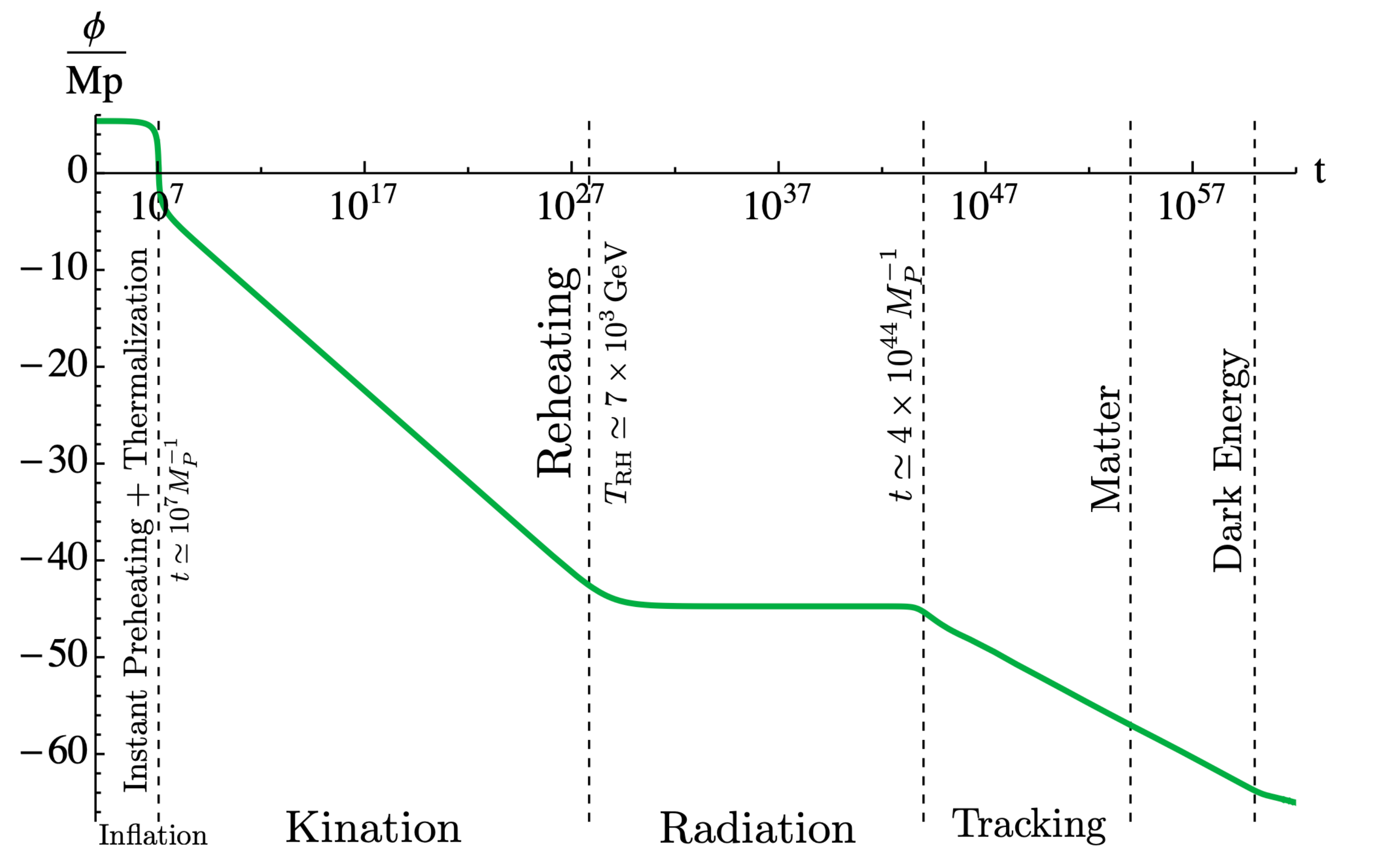}
    \caption{\it Evolution of the NO-NO inflaton field $\phi$ as a function of time in units of $M_P^{-1}$. During inflation, which lasts until $t \simeq 10^7 \, M_P^{-1}$, the inflaton 
    field value is relatively constant. When kination begins, the inflaton quickly moves to large negative values until the universe becomes radiation-dominated, and the inflaton effectively ceases to evolve when $\phi \simeq -45 \, M_P$. When the tracking regime begins,
    the inflaton starts moving again and at present  $\phi \simeq -62 \, M_P$. 
   }
    \label{fig:phivst}
\end{figure*}

The value of $\phi$ at each evolutionary stage is indicated
on the upper horizontal axis in Fig.~\ref{fig:hist}. As $\phi(t)$ is highly non-linear,
we show an extended version of Fig.~\ref{fig:plot1} in Fig.~\ref{fig:phivst}.
There we see, for example, that during inflation the inflaton field value is relatively
constant until the end of inflation (as seen in Fig.~\ref{fig:plot1}) and then drops 
rapidly, moving to negative values during the period of kination. 

\section{Preheating}
\label{sec:preheat}

Because of its coupling (\ref{pot3}, \ref{pot4}) to the inflaton field, $\phi$, the effective mass of the Higgs-like boson, $h$, starts changing non-adiabatically when the inflaton field passes near the point $\phi \simeq 0$, removing a small fraction of the inflaton kinetic energy and producing a large quantity of heavy Higgs particles through the instant preheating mechanism~\cite{instpreh}. This mechanism is extremely efficient and, for non-oscillatory models of inflation, the Universe reheats without inflaton decay and is instead reheated by the annihilations and scatterings of the effective heavy Higgs boson whose mass is $m_{\rm{h, eff}} \simeq \frac{m}{\sqrt{3}} \simeq 5.5 \times 10^{12} \, \rm{GeV}$, via the usual Yukawa interaction $\lambda_\psi h \bar{\psi} \psi$, where $\psi$ represents Standard Model fermionic fields. These massive Higgs particles
would scatter mainly into third-generation fermions, with a rate $\propto n_h/m_h^2 \gg H$, as will be seen below when we derive
and expression for the Higgs number density, $n_h$.

After the rapid scattering of Higgs bosons, the energy density of ultra-relativistic particles scales as radiation (the green curve in Fig.~\ref{fig:hist}), which scales as $\rho_r \sim a^{-4}$, and the inflaton field energy density is dominated by its kinetic energy contribution (the red curve),
which scales as $\rho_{\phi} \sim a^{-6}$. We define the reheating temperature, $T_{\rm{RH}}$, as the temperature when $\rho_r \simeq \rho_{\phi}$, and we assume that the radiation products are thermalized by that point. As we show later, reheating happens significantly before Big Bang Nucleosynthesis, $T_{\rm{RH}} \gg 1 \, \rm{MeV}$.

\if 
In order to analyze the preheating mechanism we first calculate the effective mass of the Higgs boson. From the canonically-normalized potential~(\ref{pot3}), we find that it is given by

where the $\mu$-term $\simeq 1 \, \rm{TeV}$ is associated with the electroweak scale and $m \simeq 4 \times 10^{-6} \, M_p$ is the inflationary scale in the case with $N_* = 55$ that we are considering. When most of the Higgs particles are produced through instant preheating, $\phi \simeq 0$, and the effective Higgs mass~(\ref{higgseff}) can be approximated by
\begin{equation}
    m_{\rm{h, eff}} \simeq \sqrt{\frac{m^2}{3} + \mu^2} \simeq \frac{m}{\sqrt{3}} \, , 
\end{equation}
because $m \gg \mu$.
\fi

 It is important to note that the effective Higgs mass decreases rapidly as the inflaton field value decreases. Examining Eqs.~(\ref{pot3}) and (\ref{pot4}), we find that the $m$-dependent term becomes subdominant when $\phi \lesssim -11.9 \, M_P$. This happens significantly before electroweak symmetry breaking.
 Therefore, the first term is subdominant at the present day and makes a negligible contribution to the physical Higgs boson mass.

The equation of motion for the Higgs field $h$ in Fourier space is
\begin{equation}
\label{eompre1}
    \ddot{h}_k + 3 \frac{\dot{a}}{a} \dot{h}_k + \left( \frac{k^2}{a^2} + m_{\rm{h, eff}}^2 \right) h_k = 0 \, ,
\end{equation}
where $k$ is its momentum, and the angular frequency $\omega_k(t)$ is defined as
\begin{equation}
\label{angfrq}
    \omega_k^2(t) = \frac{k^2}{a(t)^2} + m_{\rm{h, eff}}^2(t) \, .
\end{equation}
When considering a broad parametric resonance in an expanding Universe, it is convenient to introduce a new variable $H_k = a^{3/2} h_k$, which incorporates the Hubble friction term in (\ref{eompre1}), so that the equation of motion becomes
\begin{equation}
\label{modeeq}
    \ddot{H}_k + \omega_k^2 H_k = 0 \, ,
\end{equation}
where 
\begin{equation}
\label{eompre2}
    \omega_k^2 = \frac{k^2}{a(t)^2} + m_{\rm{h, eff}}^2 - \frac{3}{4} \left(\frac{\dot{a}(t)}{a(t)} \right)^2 - \frac{3}{2} \frac{\ddot{a}(t)}{a(t)} \, ,
\end{equation}
and the last two terms are responsible for gravitational particle production. In the case of instant preheating their contribution is negligible, and we can approximate the time-dependent harmonic oscillator frequency as $\omega_k^2 \simeq \frac{k^2}{a(t)^2} + m_{\rm{h, eff}}^2$. 

We take as the initial condition for the equation of motion~(\ref{modeeq}) the Bunch-Davies vacuum $H_k (t) \simeq e^{-i \omega_k t}/\sqrt{2 \omega_k}$. Since the parametric resonance occurs rapidly, we can safely ignore the expansion of the Universe and approximate the frequency as $\omega_k^2 \simeq k^2 + m_{\rm h, eff}^2$ during this period.

The harmonic-oscillator equation of motion~(\ref{modeeq}) can be used to construct the invariant particle occupation number, $n_k$, for the Fourier momentum mode $k$, which is independent of the scale factor, and is given by
\begin{equation}
\label{partocc}
   n_{k}=\frac{\omega_{k}}{2}\left(\frac{\left|\dot{H}_{k}\right|^{2}}{\omega_{k}^{2}}+\left|H_{k}\right|^{2}\right)-\frac{1}{2} \, .
\end{equation}
The dominant contribution to particle production occurs when the standard adiabaticity condition is violated, i.e.,
\begin{equation}
\label{adiab}
    \dot{\omega}_k \gtrsim \omega_k^2 \, ,
\end{equation}
and we discuss two different modes of production: when $k \simeq 0$ and $k \sim m_{\rm{h, eff}}$.\footnote{Particle production becomes exponentially suppressed when $\dot{\omega}_k < \omega_k^2$.}

If we assume that the momentum values are negligibly small, the adiabaticity violation condition~(\ref{adiab}) becomes $|\dot{m}_{\rm{h, eff}}| \gtrsim m_{\rm{h, eff}}^2$ and, using~(\ref{higgseff}), we find that
\begin{equation}
\label{adiab2}
    \frac{32 m^2 e^{4 \phi}}{3(e^{\phi} + 1)^5} \frac{|\dot{\phi}|}{m_{\rm{h, eff}}^{3/2}} \gtrsim 1 \, .
\end{equation}
From numerical approximations, we find that the adiabaticity condition~(\ref{adiab2}) is violated when $ 0.1 \, M_P \, \gtrsim \phi \gtrsim -13.8 \, M_P$. 
However, because the physical Higgs particle density is $n_h \propto \int_0^{\infty} k^2 n_k$, modes with $k \simeq 0$ make a negligible contribution to the total particle number density, and total particle production can be approximated as instantaneous.

We consider next the case when $k \sim m_{\rm{h, eff}}$. Now the adiabaticity violation condition~(\ref{adiab}) becomes
\begin{equation}
\label{adiab3}
    \frac{32 m^2 e^{4 \phi}}{3(e^{\phi} + 1)^5} \frac{|\dot{\phi}|}{(k^2 +m_{\rm{h, eff}}^2)^{\frac{3}{2}}} \gtrsim 1 \, ,
\end{equation}
where we find from numerical calculations that the dominant particle production occurs when $\phi \simeq 0$. We see from Fig.~\ref{fig:plot2} that the magnitude of the
velocity of the inflaton field $\dot{\phi}$ is maximized at the end of inflation, when $\phi \simeq 0.6 \, M_P$ and $|\dot{\phi}| \simeq 0.7 \, m \, M_P$. Near $\phi \simeq 0$, when the dominant particle production occurs, the inflaton velocity is approximately $|\dot{\phi}| \simeq 0.6 \, m \, M_P$.

\begin{figure}[ht!]
    \centering
    \includegraphics[width=0.9\textwidth]{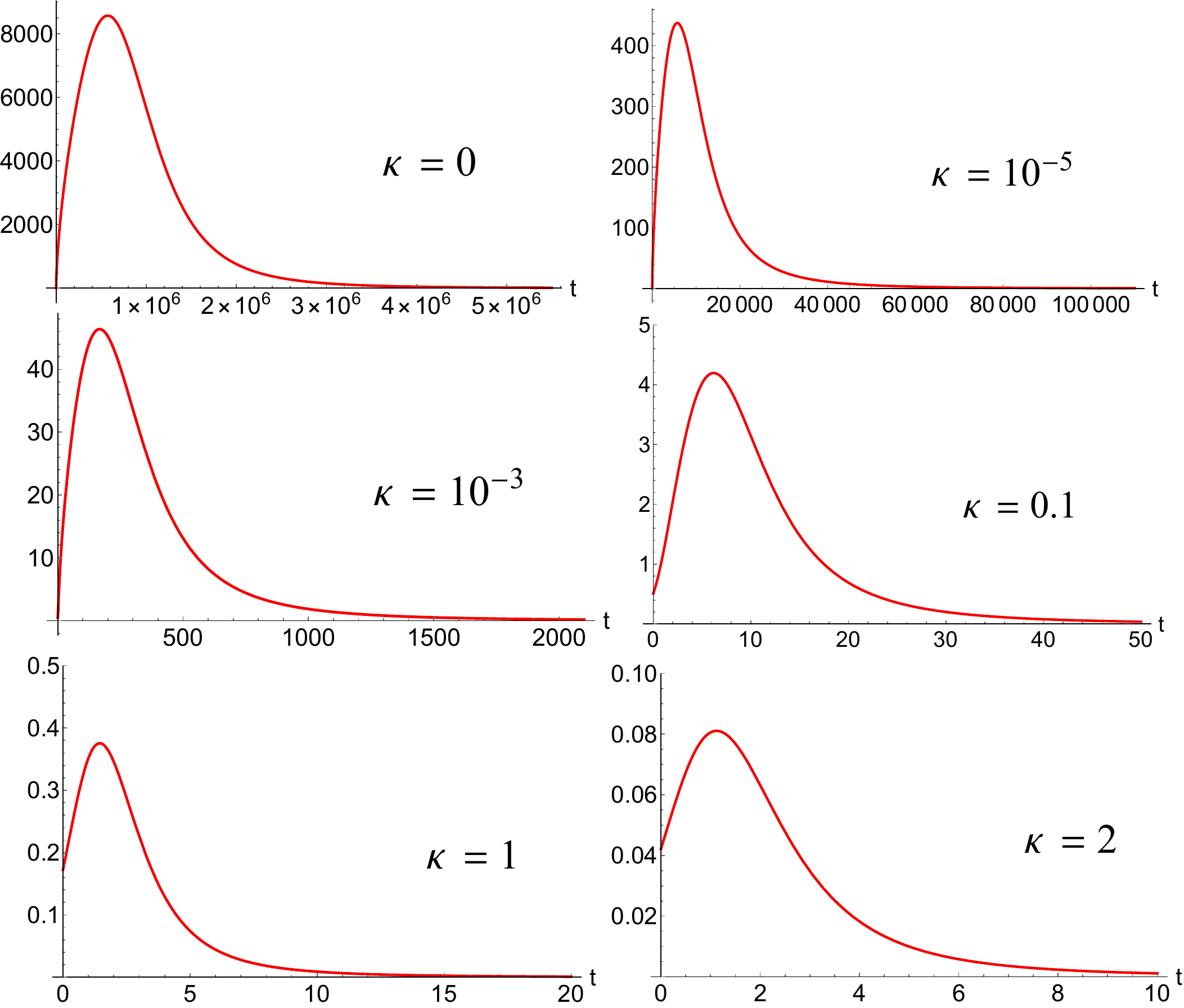}
        \caption{\it The left-hand side of the condition~(\ref{adiab3}) that corresponds to the violation of the adiabaticity condition~(\ref{adiab}) 
        as a function of time in units of $m^{-1}$ for different values of $\kappa = 0, \, 10^{-5}, \, 10^{-3}, \, 0.1, \, 1, \, \rm{and} \, 2$. For $\kappa = 1, 2$ the condition~(\ref{adiab3}) is not satisfied, and particle production is exponentially suppressed.}
    \label{fig:plot3}
\end{figure}

It should be noted that our model is quite different from the model considered in~\cite{preheating}, where the authors coupled the two boson fields through the interaction term $-\frac{1}{2} g^2 \phi^2 \chi^2$, where $\chi$ corresponds to another scalar field. In this case, the effective mass of the $\chi$ particles is given by $m_{\chi} = g |\phi|$. As the inflaton field rolls to infinity and its field value increases, the effective mass $m_{\chi}$ increases as well, until the inflaton field $\phi$ reaches its maximal value and the  $\chi$ particles decay to two fermions. In contrast, in our model the instant preheating mechanism produces heavy Higgs bosons whose effective mass $m_{\rm{h, eff}}$, given by~(\ref{higgseff}), {\em decreases} with increasing inflaton field value, which then rapidly scatter into ultra-relativistic light particles.

In order to understand better the instant preheating mechanism and adiabaticity violation condition~(\ref{adiab3}), it is convenient to introduce a dimensionless parameter $\kappa \equiv \frac{k}{m_{\rm{h, eff}}}$. We illustrate in Fig.~\ref{fig:plot3} the violation of the adiabaticity condition~(\ref{adiab3}) as a function of time for different values of $\kappa$. For these numerical calculations, we set the parameters at the end of inflation to $\phi \simeq 0.6 \, M_P, \, |\dot{\phi}| \simeq 0.7 \, m \, M_P, \, t=0, \, {\rm{and}} \, a(0) = 1$. Therefore, the condition is violated for a relatively long time when $\kappa \simeq 0$, but particle 
production is relatively instantaneous when $\kappa \sim \mathcal{O}(1)$.

\begin{figure}[ht!]
    \centering
    \includegraphics[width=\textwidth]{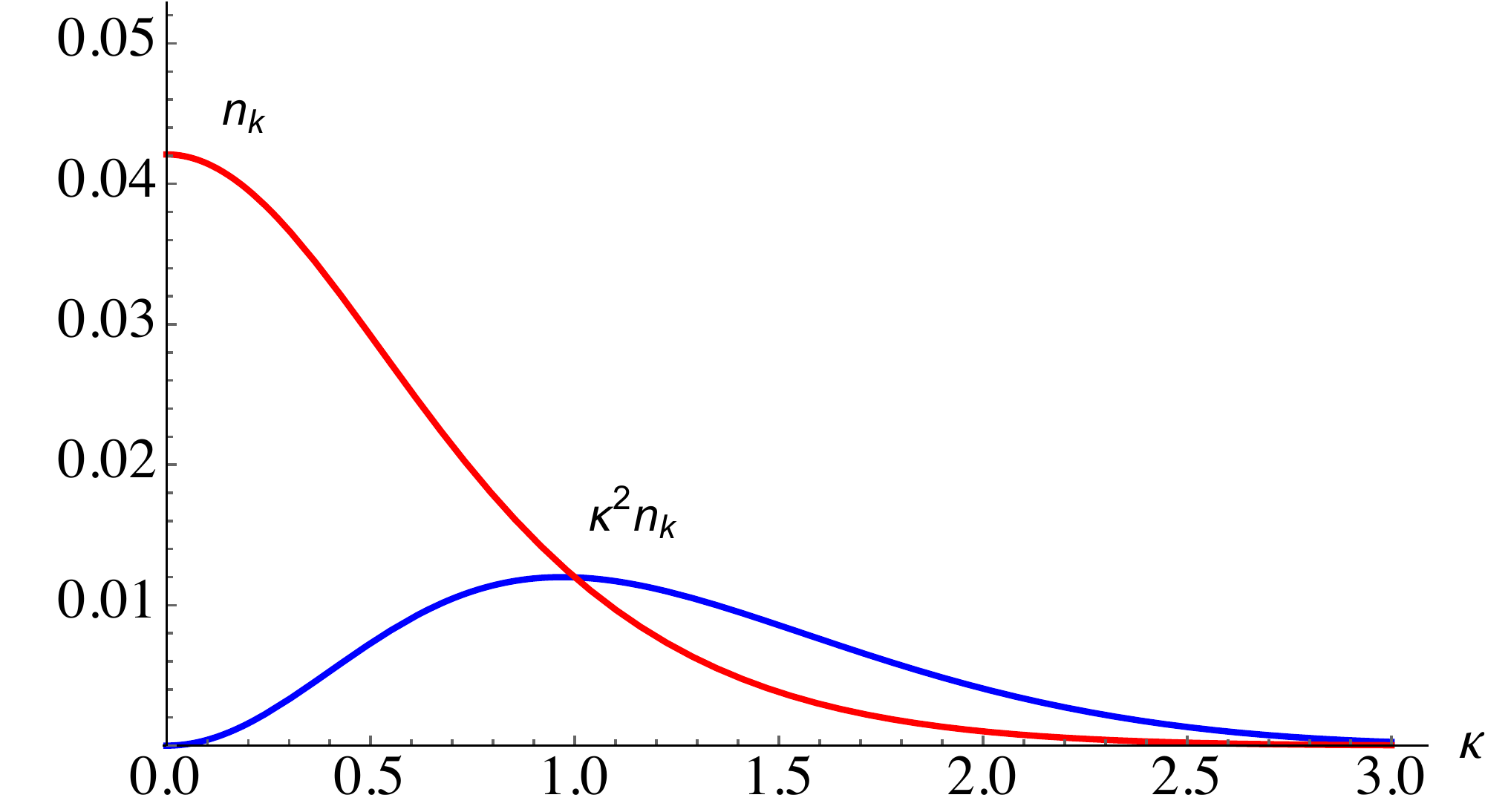}
    \caption{\it The particle occupation number $n_k$ (red) and $\kappa^2 n_k$ (blue) in units of $m^3$ as functions of $\kappa = k/m_{\rm{h, eff}}$.}
    \label{fig:plot4}
\end{figure}

In order to calculate numerically the produced Higgs boson density, we use the following expression
\begin{equation}
    n_h = \frac{1}{2 \pi^2} \int_0^{\infty} dk  \, k^2 \, n_k \, .
\end{equation}
In Fig.~\ref{fig:plot4} we show the particle occupation number $n_k$ and the integrand $k^2 n_k$ as a function of $\kappa$. We see that the dominant Higgs particle production occurs in the range $0 \lesssim k \lesssim 2$, and particle production peaks for the momentum modes $k \simeq m_{\rm{h, eff}}$.

We find from numerical calculations that the Higgs boson number density after instant preheating, when $\phi \simeq -0.5 \, M_P$ and $|\dot{\phi}| \simeq 0.5 \, m \, M_p$, is
\begin{equation}
    n_h \simeq 4 \times 10^{-4} \, m^3 \simeq 3.5 \times 10^{35} \, \rm{GeV}^3 \, ,
    \label{nh}
\end{equation}
where at $\phi \simeq - 0.5 \, M_P$, $\Gamma_{sc} \sim n_h/m_{h,{\rm eff}}^2 > H$, and the newly-created Higgs bosons thermalize.
Because the Higgs fields are produced at low momentum and are very heavy initially, we can treat them as non-relativistic particles with an energy density given by
\begin{equation}
    \rho_h \simeq n_h \, m_{\rm{h, eff}}  \simeq 1.3 \times 10^{-4} \, m^{4} \simeq 1.1 \times 10^{48} \, \rm{GeV}^4 \, ,
\end{equation}
where we use $m_{h,{\rm eff}} \simeq m/3$ from Eq.~(\ref{higgseff}) for $\phi \simeq -0.5 \, M_P$.
This is a small fraction ($\sim 10^{-15}$) of the energy density in the inflaton potential and kinetic energy which is of order $m^2 M_P^2$.
Shortly after they are produced, the heavy Higgs particles annihilate and scatter into ultra-relativistic products, $r$, and $\rho_h \simeq \rho_r$, which scales as $\propto a^{-4}$. Note that although instant preheating occurs at a time similar to the onset of kination, the Universe still expands quite rapidly before  $a \sim t^{1/3}$ becomes a good approximation which only occurs at a time $t\sim 10^8 \, M_P^{-1}$, and a significant amount of dilution occurs as seen by the dips in the energy densities shown in Fig.~\ref{fig:hist}
at $\phi \simeq 0$. This will result in a lower reheating temperature than was found in many models in the literature, as discussed in the next Section. 

We need to check that back-reaction effects are negligible during preheating, and do not affect the particle production. In order to take into account the back-reaction effects of $h$ quantum fluctuations, we use the Hartree approximation~\cite{preheating, NOinfl}, and write the equation of motion for $\phi$ as
\begin{equation}
\label{eomback}
    \ddot{\phi} + 3H \dot{\phi} + \frac{64m^2 e^{4 \phi}}{3(e^{\phi} + 1)^5} + \frac{32m^2 e^{4 \phi}}{3(e^{\phi} + 1)^5} \langle h \rangle^2 = 0 \, ,
\end{equation}
where the vacuum expectation value for $h^2$ is given by
\begin{equation}
    \left\langle h^{2}\right\rangle=\frac{1}{2 \pi^{2} a^{3}} \int_{0}^{\infty} \frac{k^{2} n_k}{\omega} dk \, .
\end{equation}
Using the angular frequency~(\ref{angfrq}), numerical calculations show that
\begin{equation}
\label{numh}
    \langle h \rangle^2 \simeq 5.6 \times 10^{-4} \left( \frac{m}{M_P} \right)^{2} \simeq 9 \times 10^{-15} \, .
\end{equation}
When instant preheating occurs we have $\phi \simeq 0$, and the equation of motion~(\ref{eomback}) becomes
\begin{equation}
\label{eomback2}
    \ddot{\phi} + 3H \dot{\phi} + \frac{2}{3}m^2 + \frac{m^2}{3} \langle h \rangle^2 = 0 \, .
\end{equation}
One can use the approximation that the back-reaction effects are negligible if the effective inflationary scale $m_{\phi, \, \rm{eff}} = m^2 + \frac{m^2}{2} \langle h \rangle^2$ does not change significantly and the term in (\ref{eomback2}) arising from the quantum fluctuations of $h$ is significantly smaller than the first term. We conclude from Eq.~(\ref{numh}) that back-reaction effects are negligible in our model.

\section{Reheating and the Electroweak Phase Transition}

The inflaton field loses only a small fraction of its energy while producing the Higgs bosons non-perturbatively. At the end of inflation, the inflaton field has an energy density $\rho_{\phi} \simeq 1.1 \times 10^{62} \, \text{GeV}^4$, and we find the following energy density ratio
\begin{equation}
\label{enratio}
    \Delta(t_{\rm{end}}) \equiv \frac{\rho_{h}(t_{\rm{end}})}{\rho_{\phi}(t_{\rm{end}})} \simeq 2 \times 10^{-15} \, ,
\end{equation}
where $t_{\rm{end}} \simeq 1.1 \times 10^7 \, M_P^{-1}$ is the time at the end of inflation.
When inflation ends, the inflaton potential energy density still dominates the total energy density until the time $t_i \simeq 1.2 \times 10^7 \, M_P^{-1}$ when kination begins (see Fig.~\ref{fig:hist}).
Between $t_{\rm end}$ and $t_i$, the Universe still expands 
rather rapidly and $\phi$ evolves from $\sim 0.6 \, M_P$ to $-0.5 \, M_P$. As noted earlier, 
even though the inflaton kinetic energy dominates the total energy density, the Universe continues to expand rapidly until $t\sim 10^8 \, M_P^{-1}$, when $H = 1/3t$ becomes a good approximation. 

The radiation bath is quickly formed after preheating.
The Higgs bosons annihilate and scatter with a rate 
$\Gamma_{sc} \sim n_h/m_{h,{\rm eff}}^2$.
As the inflaton evolves to more negative field values,
the effective mass of the Higgs boson drops rapidly
and the scattering rate quickly overtakes the expansion rate,
leading to the formation of the thermal bath. We estimate that this occurs at
$t_{sc} \sim 1.25 \times 10^7 \, M_P^{-1}$ when $\phi \simeq -1.65 \, M_P$. The energy density in radiation at this time is 
$\rho_h \simeq 3.6 \times 10^{46} \, \rm{GeV}^4$ having been diluted by a factor of $\sim 5$ relative to the naive estimate which assumes $H \sim 1/3t$ at $t_i$. This
corresponds to a maximum temperature
\beq
\label{Tmax}
T_{\rm max} = \left(\frac{30 \rho_h(t_{\rm end})}{\pi^2 g_*}\right)^{1/4} \simeq 1.5 \times 10^{11} \, {\rm GeV} \, ,
\eeq
where $g_* = 915/4$ is the number of relativistic degrees of freedom,
assuming instantaneous thermalization at $t_{sc}$. The temperature, however, quickly drops (by a factor of $\sim 30$ over that due to kination)
as the Universe begins its kination phase as seen in Fig.~\ref{fig:hist}. 

The thermal bath does
not dominate the energy density of the Universe when it is formed, and initially the radiation density drops faster than $\rho_\phi$. At $t_{sc}$, $\Delta$ is similar to that in Eq.~(\ref{enratio}).  However,  
during kination, $\rho_{\phi} \propto a^{-6}$ and $\rho_r \propto a^{-4}$. 
While the Universe is dominated by the kinetic energy of the inflaton, we have $a \sim t^{1/3}$, and $\Delta \sim a^{2} \sim t^{2/3}$. 

We define the reheating temperature $T_{\rm{RH}}$ as when $\rho_{\phi} = \rho_{r}$, so that reheating occurs when $\Delta(t_{\rm{RH}}) = 1$, which occurs when $t_{\rm{RH}} \simeq 7 \times 10^{27} \, M_P^{-1}$. The radiation energy density at reheating is 
\begin{eqnarray}
    & \rho_r(t_{\rm{RH}})& \, \simeq \, \rho_r(t_{sc}) \left(\frac{t_{sc}}{t_{\rm{RH}}} \right)^{4/3} \simeq \; 1.8 \times 10^{17} \, \text{GeV}^4
\end{eqnarray}
according to our numerical calculations.
This corresponds to a reheating temperature of
\begin{equation}
    T_{\rm RH} = \left(\frac{30 \rho_r(t_{\rm{RH}})}{\pi^2 g_*} \right)^{1/4} \simeq 7 \, \rm {TeV} \, . 
    \label{reheating}
\end{equation}

During the radiation-dominated period,
the inflaton kinetic energy decreases rapidly,
while the inflaton field value and its potential remain roughly constant,
as seen in Figs.~\ref{fig:hist} and \ref{fig:phivst}.
Radiation domination continues until $\rho_r$ (which drops as $a^{-4}$) is similar to $\rho_\phi$, whereupon the Universe enters a tracking regime~\cite{Steinhardt:1999nw}. It is easy to verify that, 
for the inflaton potential (\ref{pot4}),
the two tracking criteria
\beq
\Gamma \equiv \frac{V''V}{(V')^2} \ge 1 ~~{\rm and}~~ \left|\frac{d\ln(\Gamma - 1)}{d \ln a}
\right| \ll 1 , 
\label{tc} 
\eeq 
are both satisfied, since $\Gamma = 1 - \frac{e^\phi}{4}$ and is very nearly constant when $\phi \lesssim -8~M_P$.
Tracking begins when $t \simeq 10^{44} \, M_P^{-1}$,
corresponding to a temperature $T = 70$ keV. 
During tracking, the radiation energy density and the kinetic and potential energy densities of $\phi$ 
track each other, as seen in Fig.~\ref{fig:hist}.
As a result, the inflaton field value evolves as well, 
as seen in Fig.~\ref{fig:phivst}. 
At this time, the energy density in the inflaton is about 
1/3 the total energy density. While tracking begins after BBN has begun and therefore does not affect weak interaction freeze-out or the deuterium bottleneck, there may be some residual effects on BBN that merit further study. 

Tracking continues until the cold dark matter density begins to dominate the energy density at $t \simeq 10^{54} \, M_P^{-1}$,
and shortly thereafter vacuum energy density takes over.
Tuning is required to give the present observed values 
of the fractions of the critical density in dark energy, matter, and radiation $\Omega_\Lambda, \, \Omega_m$, and $\Omega_r$, respectively.

Before concluding this Section, we comment on the probable effect of including additional degrees of freedom on reheating. In all of the preceding discussion, 
we have restricted our attention
to a simplified model with three fields, $x, y, z$. 
We assumed that $y$ is a proxy for the Higgs field,
and its coupling to the inflaton field $x$ in Eq.~(\ref{pot3}) 
was derived from an expansion of the K\"ahler potential.
In a more realistic model, there would be $N \sim 50$ chiral superfields 
with similar couplings. We expect Eq.~(\ref{nh})
to hold for each of these, and therefore the
total number density of fields would scale with $N$.
Since the scattering rate would also scale as $N$, $\Gamma_{sc} \sim N^2 \, n_{h}/m_{\rm{h, eff}}^2 > H(t_{\rm{end}})$, thermalization
would occur immediately after preheating, and we would expect a higher 
maximal temperature $T_{\rm max} \simeq  10^{12}$ GeV. 
The increased radiation density would also lead to a higher reheating temperature, 
which we estimate would be roughly
$T_{\rm RH} \sim 4 \times 10^5$ GeV.
In this case, tracking would begin earlier,
and constraints from BBN may become relevant. However, in this paper we present only the model with a single field associated with the Higgs field, and leave a complete treatment in the minimal supersymmetric extension of the SM (MSSM)  for study in future work.

\section{Gravitational Wave Spectrum}
\label{sec:GWs}

The fractional energy density of primordial gravitational waves (GWs)
relative to the critical energy density in a flat Universe is given by~\cite{grav1}
\begin{equation}
\label{grav1}
    \Omega_{\rm{GW}} = \frac{1}{\rho_c} \frac{d \, \rho_{\rm{GW}}}{d \ln k}  \propto k^{2 \left(\frac{3w - 1}{3w + 1} \right)} \propto a^{3w -1} \, ,
\end{equation}
where $\rho_c \equiv 3H_0^2 M_{P}^2 \simeq 1.05 \, h^2 \times 10^{-5} \, {\rm{GeV}} \, {\rm{cm^{-3}}} \, $
is the critical density of the Universe, $\rho_{\rm{GW}}$ is the energy density of the gravitational waves, $k = aH$ is the momentum mode at the Hubble horizon crossing, and $w = p/\rho$ is the effective equation of state parameter. When the Universe becomes dominated by the radiation density, $w = 1/3$, and the gravitational wave spectrum remains flat because $\Omega_{\rm{GW}} \sim \rm{const.}$

However, during the kination epoch, $w = 1$, and the fractional energy density of gravitational waves scales as $\Omega_{\rm{GW}} \sim k \sim a^2$. Therefore, a prolonged period of the inflaton kinetic energy regime could lead to a blue-tilted gravitational wave spectrum capable of contributing significantly to the total radiation density of the Universe and, in turn, affecting the effective number of neutrino species $N_{\rm{eff}}$ at the time of BBN.

One can use the current limit on 
$N_{\rm eff} < 3.17$ \cite{foy2} to set an upper limit on the energy density of gravitational waves today~\cite{maggiore},
\beq
\Omega_{\rm GW} < \left(\frac{4}{11} \right)^{4/3} \frac78 (N_{\rm eff} -3) \Omega_\gamma \, ,
\eeq
where $\Omega_\gamma = (\pi^2/15) T_0^4/\rho_c \simeq 2.47 \times 10^{-5} \, h^{-2}$
and $T_0~=~2.73$ K is the present temperature of the microwave background. 
In order to ensure that higher-frequency gravitational waves do not affect
BBN, we impose the bound~\cite{maggiore}
\begin{equation}
\label{bound1}
    I = h^2 \int_{k_{\rm{BBN}}}^{k_{{\rm i}}} \Omega_{\rm{GW}} \, d \ln k \lesssim 10^{-6} \, ,
\end{equation}
where $k_{\rm{BBN}}$ and $k_{i}$ are the momentum modes associated with BBN and the onset of kination, respectively. We find numerically that the dominant contribution to the integral~(\ref{bound1}) comes from the momentum modes when $0.9 \lesssim w \lesssim 1$, which is the case in the range $k_{\rm{rad}} \leq k \leq k_{\rm kin}$, where $k_{\rm{rad}}$ is the momentum mode when the radiation energy density becomes significant, and
$k_{\rm kin}$ is the momentum mode when the kinetic energy of the inflaton is significantly larger than its potential energy.~\footnote{A common approximation is that contributions to the integral~(\ref{bound1}) come from the entire range
$[k_i, k_{\rm RH}]$, where the latter corresponds to the momentum mode when radiation dominates.
However, this overestimates the contributions to the integral from the ranges $[k_i, k_{\rm kin}]$ and $[k_{\rm rad}, k_{\rm RH}]$ where $w < 0.9$, e.g., $w \simeq 0.1$ at $k_i$.}

During the kination epoch, the fractional energy density of the gravitational waves is given by~\cite{giovannini}
\begin{equation}
\label{grav2}
    \Omega_{\rm{GW}} \simeq \varepsilon \Omega_{\gamma} h_{\mathrm{GW}}^{2}\left(\frac{k}{k_{\rm{rad}}}\right)\left[\ln \left(\frac{k}{k_{\rm{kin}}}\right)\right]^{2} \, ,
\end{equation}
where $\varepsilon = 2 R_i \left(\frac{3.36}{g_{*}} \right)^{1/3}$ with $R_i = \frac{81}{32 \pi^3}$, which takes into account the contribution of the massless scalar degrees of freedom,  and $h_{\mathrm{GW}}^{2} = \frac{1}{8 \pi} \left(\frac{H(t_{\rm{end}})}{M_P} \right)^2$ is the dimensionless gravitational wave amplitude.

Using the expression~(\ref{grav2}) as the integrand in~(\ref{bound1}), we find
\begin{equation}
    I \simeq 2 \varepsilon h_{\rm{GW}}^2 \Omega_{\gamma} h^2 \left(\frac{k_{\rm{kin}}}{k_{\rm{rad}}} \right) \, ,
\end{equation}
where we have neglected the subdominant logarithmic contribution to the integral because $k_{\rm{kin}} \gg k_{\rm{rad}}$. Next, we rewrite the ratio $k_{\rm{kin}}/k_{\rm{rad}}$ as 
\begin{equation}
    \frac{k_{\rm{kin}}}{k_{\rm{rad}}} = \frac{a_{\rm{kin}}}{a_{\rm{rad}}} \frac{H_{{\rm{kin}}}}{H_{\rm{rad}}} \simeq  2.2 \times 10^{13},
\end{equation}
and we find
\begin{equation}
    I \simeq 0.3 \, \varepsilon \,  \Omega_{\gamma} \, h^2 .
\end{equation}
Inserting the numerical values, we find
\begin{equation}
    I \simeq 6 \times 10^{-7} \, ,
\end{equation}
which is consistent with the bound~(\ref{bound1}),
 and would be reduced if additional chiral superfields couple to the inflaton.

\section{Gravitino Production}
\label{sec:gravitino}

Gravitinos can be produced after inflation
either by the direct decay of the inflaton
or through thermal production in the newly-created radiation bath \cite{weinberg,elinn,nos,ehnos,kl,ekn,Juszkiewicz:gg,mmy,Kawasaki:1994af,Moroi:1995fs,enor,Giudice:1999am,bbb,cefo,kmy,stef,Pradler:2006qh,ps2,rs,kkmy,egnop,Garcia:2017tuj}. 
In more conventional oscillatory models, inflaton decay products thermalize rapidly to a temperature $T_{\rm max}$. Then, while the Universe is dominated by the inflaton oscillations and undergoes effective matter-dominated expansion, the temperature of the radiation falls as $T \sim a^{-3/8}$ until the Universe becomes radiation-dominated at the scale defined as 
$T_{\rm RH}$~\cite{Giudice:2000ex,egnop,Garcia:2017tuj,Kaneta:2019zgw,Garcia:2020eof,preUV,Bernal:2020gzm}.
In this case, the relevant Boltzmann 
equation can be written as 
\beq
\frac{dY}{dT} = -\frac83 \frac{R}{H T^9} \, ,
\eeq
where $Y \equiv n_{3/2}/T^8$,  $R = \langle \sigma v \rangle \zeta(3)^2 T^6/\pi^4$ is the production rate per unit volume, and  $ \langle \sigma v \rangle$ is the gravitino production cross section \cite{egnop,Garcia:2017tuj}.
While dominated by inflaton oscillations, the Hubble parameter may be written as $H = \sqrt{\alpha/3}T^4/T_{\rm RH}^2 M_P$.
Integrating the equations of motion from $T_{\rm max}$ to
$T_{\rm RH}$, one finds that the abundance of gravitinos is linear in the reheating temperature with a negligible dependence on $T_{\rm max}$,
\begin{align}
\frac{n_{3/2}}{s} & \simeq \frac{\zeta(3)^2}{\pi^4 \sqrt{3 \alpha^3}} \left( M_P^2 \langle \sigma v \rangle \right) \frac{T_{\rm RH}}{M_P} \nonumber \\ 
&\simeq  
1.6 \times 10^{-18} \left(1 + 0.89 \frac{m_{1/2}^2}{m_{3/2}^2} \right) \frac{T_{\rm RH}}{10^4\  {\rm GeV}} \,,
\label{n32s}
\end{align}
where $\alpha = \pi^2 g_*/30$,   $m_{1/2}$ is a universal high-scale gaugino mass and
the terms in the parentheses correspond to the production of the transverse and longitudinal gravitino components, respectively.  

However, unlike oscillatory models of inflation, in our NO-NO model reheating is not produced by inflaton decay. 
Therefore, gravitino production occurs solely through  scattering in the thermal plasma produced by Higgs decays. 
As noted earlier, while the Universe is in the epoch of kination,
a radiation bath is produced with initial temperature $T_{\rm max}$
and then cools as $T \sim a^{-1}$. In this case, the relevant Boltzmann
equation can be written as 
\beq
\label{NONOBoltz}
\frac{dY}{dT} = -\frac{R}{H T^4} \, ,
\eeq
where now $Y \equiv n_{3/2}/T^3$, and the Hubble parameter may be written as $H = \sqrt{\alpha/3}T^3/T_{\rm RH} M_P$ during the kination-dominated period.
Integration of (\ref{NONOBoltz}) yields
\begin{align}
\frac{n_{3/2}}{s} & \simeq \frac{3 \sqrt{3} \zeta(3)^2}{4 \pi^4 \alpha^{3/2}} \left( M_P^2 \langle \sigma v \rangle \right) \frac{T_{\rm RH}}{M_P} \ln \frac{T_{\rm max}}{T_{\rm RH}} \nonumber \\ 
&\simeq  
3.7 \times 10^{-18} \left(1 + 0.89 \frac{m_{1/2}^2}{m_{3/2}^2} \right) \frac{T_{\rm RH}}{10^4\ {\rm GeV}} \ln \frac{T_{\rm max}}{T_{\rm RH}} \,.
\label{n32}
\end{align}
We see that in addition to a factor of $\sim 2$ enhancement of the numerical prefactor, 
there is a substantial logarithmic boost
factor $\ln T_{\rm max}/T_{\rm RH}$ due to production during the 
kination-dominated era.~\footnote{Particle production during kination was also considered in~\cite{preUV}, though the case
applicable to gravitino production was not considered there.}

We assume that the gravitino is not the lightest supersymmetric particle (LSP), in which case it
is expected to be unstable with a lifetime $\Gamma_{3/2} \sim m_{3/2}^2/M_P^2$. Since this rate
is relatively small for $m_{3/2} \ll M_P$, a light gravitino may decay so late as to disturb BBN.
We assume that the gravitino is sufficiently heavy, $\gtrsim {\cal O}(10)$~TeV, for this
not to be an issue.
However, the decays of the gravitinos eventually produce lighter supersymmetric particles, in particular the lightest neutralino,  $\chi$, and the inflatino, which is much lighter than $\chi$ in the simplified NO-NO model presented here, in which we have not discussed how supersymmetry breaking is communicated to the supersymmetric partners of SM particles. In this simplified model $\chi$ is a Higgsino with a lifetime for decays
to inflatinos that vastly exceeds the age of the Universe, in which case it is a candidate for cold dark matter. Requiring $\Omega_\chi h^2 \lesssim 0.12$~\cite{planck18} and assuming conservatively that most gravitino decays produce $\chi$ particles, we obtain an upper limit on the gravitino density relative to the entropy density:
\beq
\frac{n_{3/2}}{s} \lesssim 4.4 \times 10^{-12} \left( \frac{100 \, {\rm GeV}}{m_\chi} \right) \, .
\label{limit}
\eeq
For sufficiently heavy gravitinos ($m_{3/2} \gtrsim 10$ TeV), this bound is more important than that from BBN (see, e.g., \cite{ceflos}). We can combine 
the expected abundance from scattering in Eq.~(\ref{n32}) and the limit in Eq.~(\ref{limit}) to find the constraint
\begin{equation}
    \Omega_{\mathrm{DM}} h^{2} \simeq 10^{-7}\left(\frac{m_{\chi}}{100 \, \mathrm{GeV}}\right)\left(\frac{T_{\rm{RH}}}{10^{4}   \, \mathrm{GeV}}\right) \ln \frac{T_{\rm max}}{T_{\rm RH}} \lesssim 0.12 \, ,
    \label{relic}
\end{equation}
where we have assumed in Eq.~(\ref{relic}) that $m_{1/2} \ll m_{3/2}$, so that the production of the longitudinal components can be ignored. 
Using the values of $T_{\rm max}$ (\ref{Tmax}) and $T_{\rm RH}$ (\ref{reheating}) estimated earlier, 
we find $m_{\chi} \lesssim 10 \, \rm{PeV}$, which is compatible with generic supersymmetric models of dark matter.
Even in a more complete NO-NO scenario with $N\sim 50$ chiral fields,
the increased reheating temperature would still allow $m_{\chi} \lesssim 200 \, \rm{TeV}$. 
These limits do not include the additional thermal freeze-out production of neutralinos. 
Consequently, the right-hand side of (\ref{relic}) should be reduced by the value of $\Omega_\chi h^2$
left over from thermal freeze-out \cite{ehnos}, strengthening the limit on the direct production of gravitinos. 
However, these limits could be relaxed if 
there were late-time entropy production (see, e.g., \cite{ego,EGNNO23,EGNNO45,ENOV4}).

However, this discussion is over-simplified, since it neglects supersymmetry breaking in the visible sector. Moreover, it is 
also possible that the LSP is in some hidden sector~\cite{Hidden}, or that $R$-parity is broken.
We plan to return to these issues in a future paper, but conclude provisionally that LSP dark matter production in gravitino decays is not a showstopper for the NO-NO model.

\section{Results and Discussion}
\label{sec:wrap}

We have presented in this paper a novel model of no-scale supergravity inflation in which there are no inflaton oscillations at the end of the inflationary era. Instead, the inflaton preheats instantly through a coupling to a Higgs-like field that subsequently decays into relativistic matter particles. The density of the Universe is then dominated by the inflaton kinetic energy density (kination) until the density of ultra-relativistic particles starts to dominate. Reheating occurs at a temperature ${\cal O}(10^3 - 10^5)$~GeV, far above that of Big Bang Nucleosynthesis. Subsequently, the inflaton field then rolls to infinity while losing energy density.
This continues until the radiation density falls to a level near the inflaton potential energy, and tracking begins where the inflaton kinetic and potential energy density are similar to that of radiation. Eventually 
the Universe is dominated by cold dark matter, and finally dark energy. 

We have estimated the production of gravitational waves and gravitinos in this non-oscillatory scenario. We have found that the density of relic gravitational waves is below the present upper limit on the effective number of light neutral particle species imposed by the success of conventional BBN calculations. We
have also estimated
the density of LSPs produced by gravitino decays, finding that this is 
below the upper limit on the dark matter density from
Planck~2018 data~\cite{planck18} throughout the expected range of LSP masses.
One issue that we have not addressed in this paper is baryogenesis, but we
anticipate that leptogenesis~\cite{FY}, the Affleck-Dine mechanism~\cite{AD} and electroweak-scale
baryogenesis should all be possible in this NO-NO scenario.

We conclude that this no-oscillation model that invokes preheating provides an interesting 
alternative to no-scale supergravity models in which inflaton oscillations reheat the Universe,
and warrants further study.

\subsection*{Acknowledgements}

\noindent
The authors thank Marcos Garc{\' i}a for useful discussions. The work of JE was supported in part by the United Kingdom STFC Grant ST/P000258/1, and in part by the Estonian Research Council via a Mobilitas Pluss grant. The work of DVN was supported in part by the DOE grant DE-FG02-13ER42020 at Texas A\&M University and in part by the Alexander S. Onassis Public Benefit Foundation. The work of KAO was supported in part by the DOE grant DE-SC0011842 at the University of Minnesota.

\appendix
\numberwithin{equation}{section}
\renewcommand{\theequation}{\thesection\arabic{equation}}
\section{K\"ahler Transformations and the Choice of Basis}
\label{sec:appA}

Field transformations relate equivalent representations of the same no-scale model in different bases. We give some examples in this Appendix.

As a first example, consider the following form for the SU(3, 1)/SU(3)$\times$U(1) K\"ahler potential
\begin{equation}
\label{kahsu3}
    K \; = \; -\alpha \log \left(T + \bar{T} - |\phi|^2 - |S|^2 \right) \, ,
\end{equation}
where the complex field $T$ is associated with the volume modulus, and the complex fields $\phi$ and $S$ are matter-like fields. In order to express 
the K\"ahler potential~(\ref{kahsu3}) in the symmetric form~\cite{EKN2, Avatars, ENOV1}
\begin{equation}
\label{kahsu3sym}
    K \; = \; - \alpha \, \log \left(1 - x \bar{x} - y \bar{y} - z \bar{z} \right) \, ,
\end{equation}
we relate the complex scalar fields $T, \, \phi$, and $S$ to the complex fields $x, y, z$ appearing in the symmetric K\"ahler potential form~(\ref{kahsu3sym}) by using the transformations
\begin{align}
\label{transA1}
  x \; & = \; \frac{1 - 2T}{1 + 2T} \, ; & y \; & = \; \frac{2 \phi}{1 + 2T};  & z \; & = \; \frac{2S}{1 + 2T} \, ,\\ \intertext{whose inverse transformations are} 
\label{transA2}
  T \; & = \; \frac{1}{2} \left( \frac{1 - x}{1 + x} \right) \, ; & \phi \; & = \; \frac{y}{1 + x} \, ; & S  \; & = \; \frac{z}{1 + x} \, .
\end{align}
When the fields are transformed according to Eqs.~(\ref{transA1}, \ref{transA2}), the effective superpotential in the $(T, \phi, S)$ basis transforms as
\begin{equation}
    W (T, \phi, S) \longrightarrow \widetilde{W} (x, y, z) = \left(1 + x \right)^{\alpha} W(T, \phi, S) \, ,
\end{equation}
which leaves invariant the extended K\"ahler potential $G$.

Similarly, the K\"ahler potential that parametrizes an (SU(2, 1)/SU(2)$\times$U(1))$\times$(SU(1, 1)/U(1) coset manifold
\begin{equation}
\label{kahsu2su1}
    K \; = \; -\alpha \log \left(S + \bar{S} - |\phi|^2\right) - \beta \log \left(T + \bar{T} \right) \, ,
\end{equation}
can be transformed to a symmetric form for the K\"ahler potential by applying the following transformations
\begin{align}
\label{transA3}
  x \; & = \; \frac{1 - 2S}{1 + 2S} \, ; & y \; & = \;\frac{2 \phi}{1 + 2S};  & z \; & = \; \frac{1 - 2T}{1 + 2T} \, ,\\ \intertext{whose inverse relations are} 
\label{transA4}
  T \; & = \; \frac{1}{2} \left( \frac{1 - z}{1 + z} \right) \, ; & \phi \; & = \; \frac{y}{1 + x} \, ; & S  \; & = \; \frac{1}{2} \left( \frac{1 - x}{1 + x} \right) \, .
\end{align}
If we transform the coordinates using the relations~(\ref{transA3})-(\ref{transA4}), the effective superpotential transforms as
\begin{equation}
    \label{suptransf}
    W(T, \phi, S) \longrightarrow \widetilde{W} (x, y, z) = \left(1 + x \right)^{\alpha} \left(1 + z \right)^{\beta} W(T, \phi, S) \, ,
\end{equation}
again leaving invariant the extended K\"ahler potential $G$.


\section{Supersymmetry Breaking}

In this Appendix we follow the treatment presented in~\cite{ENOV2, ENOV3, ENOV4} and introduce an adjustable parameter for supersymmetry breaking without invoking additional fields. We introduce the following superpotential in the $(T, \phi, S)$ basis:
\begin{equation}
    W (T, \phi, S) = W_{\rm{inf}} + W_{\rm{SUSY}} ~, ,
\end{equation}
where $W_{\rm{inf}}$ is the inflationary superpotential part and $W_{\rm{SUSY}}$ is responsible for supersymmetry breaking, and is given by
\begin{equation}
    W_{\rm{SUSY}} (T, \phi, S) \; = \; \lambda \left(2S - \frac{\phi^2}{3} \right)^{\frac{1}{2}(\alpha + \sqrt{3 \alpha})}  \left( 2T \right)^{\beta/2} \, ,
\end{equation}
where $\lambda$ is an adjustable constant. It is important to note that the supersymmetry-breaking superpotential $W_{\rm{SUSY}}$ does not affect the inflationary potential and also does not spoil its non-oscillatory behavior.

If we use the transformations~(\ref{transA4}) together with the superpotential transformation~(\ref{suptransf}), we find that the supersymmetry-breaking superpotential contribution in the symmetric $(x, y, z)$ basis is given by
\begin{equation}
    \label{susybreak}
    W_{\rm{SUSY}} (x, y, z) = \lambda \, \left(1 - x^2 - y^2 \right)^{\frac{1}{2}(\alpha + \sqrt{3 \alpha})} \left(1 + x \right)^{-\sqrt{3 \alpha}}  \left(1 - z^2 \right)^{\beta/2} \, ,
\end{equation}
where we dropped the tilde over the left-hand side. In this case, supersymmetry breaking is generated through an $F$-term, which is given by
\begin{equation}
    F^i = -e^{G/2} K^{i \bar{j}} G_{\bar{j}} \, .
\end{equation}
Combining the superpotential~(\ref{susybreak}) with the inflationary superpotential~(\ref{sup1}) and setting the model parameters to $c = 4$, $b = \alpha = 2$, and $\beta = 1$, we find that supersymmetry is broken through an $F$-term for the inflaton field, $x$, which is given by
\begin{equation}
    |F^{x}|^2 \; = \; \frac{6 \lambda^2 e^{-\sqrt{6} \phi}}{\left(1 + \cosh{\phi} \right)^2} \neq 0 \, , \quad {\rm{when}} \quad \lambda \neq 0 \, ,
\end{equation}
where we used the canonical normalization of the field $x$ and safely ignored the contribution arising from the field $\langle y \rangle \ll \langle x \rangle$, which is associated with the Higgs field. We find numerically that supersymmetry breaking at the present day, when $\phi \simeq - 62$, is given by
\begin{equation}
    |F_{x}|^2 \; \simeq \; 3 \times 10^{13} \, \lambda^2 \, .
\end{equation}




\begin{thebibliography}{9}

 \bibitem{reviews}
   K.~A.~Olive,
  Phys.\ Rept.\  {\bf 190} (1990) 307;
A. D. Linde, {\it Particle  
Physics and
Inflationary Cosmology} (Harwood, Chur, Switzerland, 1990); 
  D.~H.~Lyth and A.~Riotto,
{\it Phys.\ Rep.}  {\bf 314} (1999) 1
[arXiv:hep-ph/9807278];
A.~D.~Linde,
Phys. Rept. \textbf{333}, 575-591 (2000);
J.~Martin, C.~Ringeval and V.~Vennin,
  Phys.\ Dark Univ.\  {\bf 5-6}, 75-235 (2014)
  [arXiv:1303.3787 [astro-ph.CO]];
  J.~Martin, C.~Ringeval, R.~Trotta and V.~Vennin,
  JCAP {\bf 1403} (2014) 039
  [arXiv:1312.3529 [astro-ph.CO]];
 J.~Martin,
  Astrophys.\ Space Sci.\ Proc.\  {\bf 45}, 41 (2016)
  [arXiv:1502.05733 [astro-ph.CO]].
  

  
\bibitem{reheating}  
A.~Dolgov and A.~D.~Linde,
Phys. Lett. B \textbf{116}, 329 (1982);
L.~Abbott, E.~Farhi and M.~B.~Wise,
Phys. Lett. B \textbf{117}, 29 (1982);

\bibitem{nos}
  D.~V.~Nanopoulos, K.~A.~Olive and M.~Srednicki,
  Phys.\ Lett.\ B {\bf 127}, 30 (1983).
  
\bibitem{preheating} 
L.~Kofman, A.~D.~Linde and A.~A.~Starobinsky,
Phys. Rev. Lett. \textbf{73}, 3195-3198 (1994)
[arXiv:hep-th/9405187 [hep-th]];
L.~Kofman, A.~D.~Linde and A.~A.~Starobinsky,
Phys. Rev. D \textbf{56}, 3258-3295 (1997)
[arXiv:hep-ph/9704452 [hep-ph]].

\bibitem{stb}
Y.~Shtanov, J.~H.~Traschen and R.~H.~Brandenberger,
Phys. Rev. D \textbf{51}, 5438-5455 (1995)
[arXiv:hep-ph/9407247 [hep-ph]].

\bibitem{kt}
S.~Y.~Khlebnikov and I.~I.~Tkachev,
Phys. Rev. Lett. \textbf{77}, 219-222 (1996)
[arXiv:hep-ph/9603378 [hep-ph]];
S.~Y.~Khlebnikov and I.~I.~Tkachev,
Phys. Lett. B \textbf{390}, 80-86 (1997)
[arXiv:hep-ph/9608458 [hep-ph]];
S.~Y.~Khlebnikov and I.~I.~Tkachev,
Phys. Rev. Lett. \textbf{79}, 1607-1610 (1997)
[arXiv:hep-ph/9610477 [hep-ph]].

\bibitem{Greene:1997fu}
P.~B.~Greene, L.~Kofman, A.~D.~Linde and A.~A.~Starobinsky,
Phys. Rev. D \textbf{56}, 6175-6192 (1997)
[arXiv:hep-ph/9705347 [hep-ph]].

\bibitem{prerev}
B.~A.~Bassett, S.~Tsujikawa and D.~Wands,
Rev. Mod. Phys. \textbf{78}, 537-589 (2006)
[arXiv:astro-ph/0507632 [astro-ph]];
R.~Allahverdi, R.~Brandenberger, F.~Y.~Cyr-Racine and A.~Mazumdar,
Ann. Rev. Nucl. Part. Sci. \textbf{60}, 27-51 (2010)
[arXiv:1001.2600 [hep-th]];
M.~A.~Amin, M.~P.~Hertzberg, D.~I.~Kaiser and J.~Karouby,
Int. J. Mod. Phys. D \textbf{24}, 1530003 (2014)
[arXiv:1410.3808 [hep-ph]].



  
\bibitem{instpreh}
G.~N.~Felder, L.~Kofman and A.~D.~Linde,
Phys. Rev. D \textbf{59}, 123523 (1999)
[arXiv:hep-ph/9812289 [hep-ph]].




\bibitem{NOinfl}
G.~N.~Felder, L.~Kofman and A.~D.~Linde,
Phys. Rev. D \textbf{60}, 103505 (1999)
[arXiv:hep-ph/9903350 [hep-ph]].
 
\bibitem{gravprod}
L.~Parker,
Phys. Rev. Lett. \textbf{21}, 562-564 (1968);
L.~Parker,
Phys. Rev. \textbf{183}, 1057-1068 (1969);
L.~Parker,
Phys. Rev. D \textbf{3}, 346-356 (1971);
V.~Frolov, S.~Mamaev and V.~Mostepanenko,
Phys. Lett. A \textbf{55}, 389-390 (1976);
A.~Grib, B.~Levitskii and V.~Mostepanenko,
Teor. Mat. Fiz. \textbf{19}, 59-75 (1974);
A.~Grib, S.~Mamaev and V.~Mostepanenko,
Gen. Rel. Grav. \textbf{7}, 535-547 (1976);
Y.~Zel'dovich and A.~Starobinsky,
JETP Lett. \textbf{26}, no.5, 252 (1977);
L.~Ford,
Phys. Rev. D \textbf{35}, 2955 (1987).

\bibitem{Hashiba:2018iff}
S.~Hashiba and J.~Yokoyama,
JCAP \textbf{01}, 028 (2019)
[arXiv:1809.05410 [gr-qc]].

\bibitem{Haro:2018zdb}
J.~Haro, W.~Yang and S.~Pan,
JCAP \textbf{01}, 023 (2019)
[arXiv:1811.07371 [gr-qc]].

\bibitem{curvaton1}
D.~H.~Lyth and D.~Wands,
Phys. Lett. B \textbf{524}, 5-14 (2002)
[arXiv:hep-ph/0110002 [hep-ph]];
\bibitem{curvaton2}
B.~Feng and M.~z.~Li,
Phys. Lett. B \textbf{564}, 169-174 (2003)
[arXiv:hep-ph/0212213 [hep-ph]];
A.~R.~Liddle and L.~Urena-Lopez,
Phys. Rev. D \textbf{68}, 043517 (2003)
[arXiv:astro-ph/0302054 [astro-ph]];
K.~Dimopoulos,
Phys. Rev. D \textbf{68}, 123506 (2003)
[arXiv:astro-ph/0212264 [astro-ph]].
C.~Campuzano, S.~del Campo, R.~Herrera and R.~Herrera,
Phys. Rev. D \textbf{72}, 083515 (2005)
[arXiv:hep-th/0511255 [hep-th]];
C.~Campuzano, S.~del Campo and R.~Herrera,
Phys. Lett. B \textbf{633}, 149-154 (2006)
[arXiv:gr-qc/0511128 [gr-qc]].

\bibitem{kination}
M.~Joyce,
Phys. Rev. D \textbf{55}, 1875-1878 (1997)
[arXiv:hep-ph/9606223 [hep-ph]];
M.~Joyce and T.~Prokopec,
Phys. Rev. D \textbf{57}, 6022-6049 (1998)
[arXiv:hep-ph/9709320 [hep-ph]];
P.~G.~Ferreira and M.~Joyce,
Phys. Rev. D \textbf{58}, 023503 (1998)
[arXiv:astro-ph/9711102 [astro-ph]].

\bibitem{PV}
P.~Peebles and A.~Vilenkin,
Phys. Rev. D \textbf{59}, 063505 (1999)
[arXiv:astro-ph/9810509 [astro-ph]].

\bibitem{tash}
H.~Tashiro, T.~Chiba and M.~Sasaki,
Class. Quant. Grav. \textbf{21}, 1761-1772 (2004)
[arXiv:gr-qc/0307068 [gr-qc]].

\bibitem{giovannini}
M.~Giovannini,
Phys. Rev. D \textbf{60}, 123511 (1999)
[arXiv:astro-ph/9903004 [astro-ph]];
M.~Giovannini,
Class. Quant. Grav. \textbf{16}, 2905-2913 (1999)
[arXiv:hep-ph/9903263 [hep-ph]].

\bibitem{alpha2}
K.~Dimopoulos and C.~Owen,
JCAP \textbf{06}, 027 (2017)
[arXiv:1703.00305 [gr-qc]];
K.~Dimopoulos, L.~Donaldson Wood and C.~Owen,
Phys. Rev. D \textbf{97}, no.6, 063525 (2018)
[arXiv:1712.01760 [astro-ph.CO]];
Y.~Akrami, R.~Kallosh, A.~Linde and V.~Vardanyan,
JCAP \textbf{06}, 041 (2018)
[arXiv:1712.09693 [hep-th]].

\bibitem{prepbh}
A.~M.~Green and K.~A.~Malik,
Phys. Rev. D \textbf{64}, 021301 (2001)
[arXiv:hep-ph/0008113 [hep-ph]];
B.~A.~Bassett and S.~Tsujikawa,
Phys. Rev. D \textbf{63}, 123503 (2001)
[arXiv:hep-ph/0008328 [hep-ph]];
T.~Suyama, T.~Tanaka, B.~Bassett and H.~Kudoh,
Phys. Rev. D \textbf{71}, 063507 (2005)
[arXiv:hep-ph/0410247 [hep-ph]].

\bibitem{pregb}
C.~van de Bruck, K.~Dimopoulos and C.~Longden,
Phys. Rev. D \textbf{94}, no.2, 023506 (2016)
[arXiv:1605.06350 [astro-ph.CO]];

\bibitem{preUV}
N.~Bernal, J.~Rubio and H.~Veermäe,
JCAP \textbf{06}, 047 (2020)
[arXiv:2004.13706 [hep-ph]].


  \bibitem{no-scale}
E.~Cremmer, S.~Ferrara, C.~Kounnas and D.~V.~Nanopoulos,
  Phys.\ Lett.\ B {\bf 133} (1983) 61.
  
    \bibitem{Ellis:1983sf} 
  J.~R.~Ellis, A.~B.~Lahanas, D.~V.~Nanopoulos and K.~Tamvakis,
  Phys.\ Lett.\  {\bf 134B}, 429 (1984).
 
    \bibitem{LN}
  A.~B.~Lahanas and D.~V.~Nanopoulos,
  Phys.\ Rept.\  {\bf 145} (1987) 1.
  
\bibitem{Steinhardt:1999nw}
I.~Zlatev, L.~M.~Wang and P.~J.~Steinhardt,
Phys. Rev. Lett. \textbf{82}, 896-899 (1999)
[arXiv:astro-ph/9807002 [astro-ph]];
P.~J.~Steinhardt, L.~M.~Wang and I.~Zlatev,
Phys. Rev. D \textbf{59}, 123504 (1999)
[arXiv:astro-ph/9812313 [astro-ph]].

  
  \bibitem{Witten}
E.~Witten,
  Phys.\ Lett.\  {\bf 155B} (1985) 151.

\bibitem{ENO8}
    J.~Ellis, D.~V.~Nanopoulos and K.~A.~Olive,
  Phys.\ Rev.\ D {\bf 89} (2014) 4,  043502
  [arXiv:1310.4770 [hep-ph]];
  
\bibitem{EGNO4} 
  J.~Ellis, M.~A.~G.~Garc{\' i}a, D.~V.~Nanopoulos and K.~A.~Olive,
  JCAP {\bf 1510}, 003 (2015)
  [arXiv:1503.08867 [hep-ph]].
  
\bibitem{ENOV4}
J.~Ellis, D.~V.~Nanopoulos, K.~A.~Olive and S.~Verner,
[arXiv:2004.00643 [hep-ph]].
  
  \bibitem{king2}
 S.~F.~King and E.~Perdomo,
  JHEP {\bf 1905}, 211 (2019)
  [arXiv:1903.08448 [hep-ph]].
  
  
    \bibitem{egnno1}
  J.~Ellis, M.~A.~G.~Garc{\' i}a, N.~Nagata, D.~V.~Nanopoulos and K.~A.~Olive,
  JCAP {\bf 1611}, no. 11, 018 (2016)
  [arXiv:1609.05849 [hep-ph]].
  
  
\bibitem{EGNNO23}
   J.~Ellis, M.~A.~G.~Garc{\' i}a, N.~Nagata, D.~V.~Nanopoulos and K.~A.~Olive,
  JCAP {\bf 1707}, no. 07, 006 (2017)
  [arXiv:1704.07331 [hep-ph]];
  J.~Ellis, M.~A.~G.~Garc{\' i}a, N.~Nagata, D.~V.~Nanopoulos and K.~A.~Olive,
  JCAP {\bf 1904}, no. 04, 009 (2019)
  [arXiv:1812.08184 [hep-ph]].
  
\bibitem{EGNNO45}
  J.~Ellis, M.~A.~G.~Garcia, N.~Nagata, D.~V.~Nanopoulos and K.~A.~Olive,
  Phys.\ Lett.\ B {\bf 797}, 134864 (2019)
  [arXiv:1906.08483 [hep-ph]];
 J.~Ellis, M.~A.~G.~Garcia, N.~Nagata, D.~V.~Nanopoulos and K.~A.~Olive,
  JCAP {\bf 2001}, no. 01, 035 (2020)
  [arXiv:1910.11755 [hep-ph]].
 
  \bibitem{dgmo}
E.~Dudas, T.~Gherghetta, Y.~Mambrini and K.~A.~Olive,
  Phys.\ Rev.\ D {\bf 96}, no. 11, 115032 (2017)
  [arXiv:1710.07341 [hep-ph]];
  K.~Kaneta, Y.~Mambrini, K.~A.~Olive and S.~Verner,
  Phys.\ Rev.\ D {\bf 101}, no. 1, 015002 (2020)
  [arXiv:1911.02463 [hep-ph]].
  
\bibitem{planck18}
  N.~Aghanim {\it et al.} [Planck Collaboration],
  arXiv:1807.06209 [astro-ph.CO];
  Y.~Akrami {\it et al.} [Planck Collaboration],
  arXiv:1807.06211 [astro-ph.CO].

\bibitem{rlimit}
P.~A.~R.~Ade {\it et al.} [BICEP2 and Keck Array Collaborations],
  Phys.\ Rev.\ Lett.\  {\bf 121}, 221301 (2018)
  [arXiv:1810.05216 [astro-ph.CO]].
  
\bibitem{Graziani:1988bp}
F.~Graziani and K.~A.~Olive,
Phys. Lett. B \textbf{216}, 31-36 (1989).
  
\bibitem{Staro}
A.~A.~Starobinsky,
  Phys.\ Lett.\ B {\bf 91}, 99 (1980).
  
  \bibitem{ENO6}
   J.~Ellis, D.~V.~Nanopoulos and K.~A.~Olive,
  Phys.\ Rev.\ Lett.\  {\bf 111} (2013) 111301 
  [arXiv:1305.1247 [hep-th]].
  
  \bibitem{Avatars}
   J.~Ellis, D.~V.~Nanopoulos and K.~A.~Olive,
  JCAP {\bf 1310} (2013) 009
 [arXiv:1307.3537 [hep-th]].
  
  \bibitem{ENOV1}
  J.~Ellis, D.~V.~Nanopoulos, K.~A.~Olive and S.~Verner,
  JHEP {\bf 1903} (2019) 099
  [arXiv:1812.02192 [hep-th]].
  
  
  
  \bibitem{ENOV2}
 J.~Ellis, D.~V.~Nanopoulos, K.~A.~Olive and S.~Verner,
  Phys.\ Rev.\ D {\bf 100}, no. 2, 025009 (2019)
  [arXiv:1903.05267 [hep-ph]].
  
  
    \bibitem{KLno-scale}
R.~Kallosh and A.~Linde,
  JCAP {\bf 1306}, 028 (2013)
  [arXiv:1306.3214 [hep-th]].
  
  \bibitem{FKR}
F.~Farakos, A.~Kehagias and A.~Riotto,
  Nucl.\ Phys.\ B {\bf 876}, 187 (2013)
  [arXiv:1307.1137 [hep-th]].
  
  \bibitem{FeKR}
  S.~Ferrara, A.~Kehagias and A.~Riotto,
  Fortsch.\ Phys.\  {\bf 62}, 573 (2014)
  [arXiv:1403.5531 [hep-th]];
S.~Ferrara, A.~Kehagias and A.~Riotto,
  Fortsch.\ Phys.\  {\bf 63}, 2 (2015)
  [arXiv:1405.2353 [hep-th]];
 R.~Kallosh, A.~Linde, B.~Vercnocke and W.~Chemissany,
  JCAP {\bf 1407}, 053 (2014)
  [arXiv:1403.7189 [hep-th]];
 K.~Hamaguchi, T.~Moroi and T.~Terada,
  Phys.\ Lett.\ B {\bf 733}, 305 (2014)
  [arXiv:1403.7521 [hep-ph]];
 J.~Ellis, M.~A.~G.~Garc\'ia, D.~V.~Nanopoulos and K.~A.~Olive,
  JCAP {\bf 1405}, 037 (2014)
  [arXiv:1403.7518 [hep-ph]];
J.~Ellis, M.~A.~G.~Garc{\' i}a, D.~V.~Nanopoulos and K.~A.~Olive,
  JCAP {\bf 1408}, 044 (2014)
  [arXiv:1405.0271 [hep-ph]].
  
  \bibitem{adfs}
I.~Antoniadis, E.~Dudas, S.~Ferrara and A.~Sagnotti,
  Phys.\ Lett.\ B {\bf 733}, 32 (2014)
  [arXiv:1403.3269 [hep-th]].

  
  \bibitem{king}
  M.~C.~Romao and S.~F.~King,
  JHEP {\bf 1707}, 033 (2017)
  [arXiv:1703.08333 [hep-ph]].

  
   \bibitem{eno9}
  J.~Ellis, D.~V.~Nanopoulos and K.~A.~Olive,
  Phys.\ Rev.\ D {\bf 97}, no. 4, 043530 (2018)
  [arXiv:1711.11051 [hep-th]].
   
  
  
  \bibitem{others}
  S.~Ferrara, R.~Kallosh, A.~Linde and M.~Porrati,
  Phys.\ Rev.\ D {\bf 88} (2013) 8,  085038
  [arXiv:1307.7696 [hep-th]];
  W.~Buchm\"uller, V.~Domcke and C.~Wieck,
  Phys.\ Lett.\ B {\bf 730}, 155 (2014)
  [arXiv:1309.3122 [hep-th]];
  C.~Pallis,
  JCAP {\bf 1404}, 024 (2014)
  [arXiv:1312.3623 [hep-ph]];
   C.~Pallis,
  JCAP {\bf 1408}, 057 (2014)
  [arXiv:1403.5486 [hep-ph]];
     W.~Buchmuller, E.~Dudas, L.~Heurtier and C.~Wieck,
  JHEP {\bf 1409}, 053 (2014)
  [arXiv:1407.0253 [hep-th]];
    J.~Ellis, M.~A.~G.~Garc\'ia, D.~V.~Nanopoulos and K.~A.~Olive,
  JCAP {\bf 1501}, no. 01, 010 (2015)
  [arXiv:1409.8197 [hep-ph]];
   T.~Terada, Y.~Watanabe, Y.~Yamada and J.~Yokoyama,
  JHEP {\bf 1502}, 105 (2015)
  [arXiv:1411.6746 [hep-ph]];
  W.~Buchmuller, E.~Dudas, L.~Heurtier, A.~Westphal, C.~Wieck and M.~W.~Winkler,
  JHEP {\bf 1504}, 058 (2015)
  [arXiv:1501.05812 [hep-th]];
 A.~B.~Lahanas and K.~Tamvakis,
  Phys.\ Rev.\ D {\bf 91}, no. 8, 085001 (2015)
  [arXiv:1501.06547 [hep-th]];
I.~Dalianis and F.~Farakos,
  JCAP {\bf 1507}, no. 07, 044 (2015)
  [arXiv:1502.01246 [gr-qc]].
  I.~Garg and S.~Mohanty,
  Phys.\ Lett.\ B {\bf 751}, 7 (2015)
  [arXiv:1504.07725 [hep-ph]];
  J.~Ellis, M.~A.~G.~Garc{\' i}a, D.~V.~Nanopoulos and K.~A.~Olive,
  JCAP {\bf 1507}, no. 07, 050 (2015)
  [arXiv:1505.06986 [hep-ph]];
  E.~Dudas and C.~Wieck,
  JHEP {\bf 1510}, 062 (2015)
  [arXiv:1506.01253 [hep-th]];
M.~Scalisi,
  JHEP {\bf 1512}, 134 (2015)
  [arXiv:1506.01368 [hep-th]];
  S.~Ferrara, A.~Kehagias and M.~Porrati,
  JHEP {\bf 1508}, 001 (2015)
  [arXiv:1506.01566 [hep-th]];
  J.~Ellis, M.~A.~G.~Garc{\' i}a, D.~V.~Nanopoulos and K.~A.~Olive,
  Class.\ Quant.\ Grav.\  {\bf 33}, no. 9, 094001 (2016)
  [arXiv:1507.02308 [hep-ph]];
  A.~Addazi and M.~Y.~Khlopov,
  Phys.\ Lett.\ B {\bf 766}, 17 (2017)
  [arXiv:1612.06417 [gr-qc]];
   C.~Pallis and N.~Toumbas,
 Adv.\ High Energy Phys.\  {\bf 2017}, 6759267 (2017)
  [arXiv:1612.09202 [hep-ph]];
  T.~Kobayashi, O.~Seto and T.~H.~Tatsuishi,
  PTEP {\bf 2017}, no. 12, 123B04 (2017)
  [arXiv:1703.09960 [hep-th]];
 I.~Garg and S.~Mohanty,
  Int.\ J.\ Mod.\ Phys.\ A {\bf 33}, no. 21, 1850127 (2018)
  [arXiv:1711.01979 [hep-ph]];
   W.~Ahmed and A.~Karozas,
  Phys.\ Rev.\ D {\bf 98}, no. 2, 023538 (2018)
  [arXiv:1804.04822 [hep-ph]].
   Y.~Cai, R.~Deen, B.~A.~Ovrut and A.~Purves,
  JHEP {\bf 1809}, 001 (2018)
  [arXiv:1804.07848 [hep-th]].
   S.~Khalil, A.~Moursy, A.~K.~Saha and A.~Sil,
  Phys.\ Rev.\ D {\bf 99}, no. 9, 095022 (2019)
  [arXiv:1810.06408 [hep-ph]].

  
  \bibitem{ENOV3}
J.~Ellis, D.~V.~Nanopoulos, K.~A.~Olive and S.~Verner,
  JCAP {\bf 1909}, no. 09, 040 (2019)
  [arXiv:1906.10176 [hep-th]].

\bibitem{tmodel}
R.~Kallosh and A.~Linde,
JCAP \textbf{07}, 002 (2013)
[arXiv:1306.5220 [hep-th]].

\bibitem{multitmodel}
O.~Iarygina, E.~I.~Sfakianakis, D.~G.~Wang and A.~Achúcarro,
[arXiv:2005.00528 [astro-ph.CO]].

  \bibitem{EKN3}
   J.~R.~Ellis, C.~Kounnas and D.~V.~Nanopoulos,
  Phys.\ Lett.\  {\bf 143B}, 410 (1984).

\bibitem{GomezReino:2006wv}
M.~Gomez-Reino and C.~A.~Scrucca,
JHEP \textbf{09}, 008 (2006)
[arXiv:hep-th/0606273 [hep-th]].

\bibitem{Covi:2008cn}
L.~Covi, M.~Gomez-Reino, C.~Gross, J.~Louis, G.~A.~Palma and C.~A.~Scrucca,
JHEP \textbf{08}, 055 (2008)
[arXiv:0805.3290 [hep-th]].

\bibitem{grav1}
V.~Sahni,
Phys. Rev. D \textbf{42}, 453-463 (1990)

\bibitem{foy2}
B.~D.~Fields, K.~A.~Olive, T.~H.~Yeh and C.~Young,
JCAP \textbf{03}, 010 (2020)
[arXiv:1912.01132 [astro-ph.CO]].

\bibitem{maggiore}
M.~Maggiore,
Phys. Rept. \textbf{331}, 283-367 (2000)
[arXiv:gr-qc/9909001 [gr-qc]].



 \bibitem{weinberg}
  S.~Weinberg,
  Phys.\ Rev.\ Lett.\  {\bf 48}, 1303 (1982).
  
 \bibitem{elinn}
  J.~R.~Ellis, A.~D.~Linde and D.~V.~Nanopoulos,
  Phys.\ Lett.\ B {\bf 118}, 59 (1982).


    \bibitem{ehnos}
  J.~Ellis, J.~Hagelin, D.~Nanopoulos, K.~Olive and M.~Srednicki,
                Nucl.\ Phys.\ B {\bf 238} (1984) 453.
                
\bibitem{kl}
 M.~Y.~Khlopov and A.~D.~Linde,
  Phys.\ Lett.\ B {\bf 138}, 265 (1984).
                
 \bibitem{ekn}
  J.~R.~Ellis, J.~E.~Kim and D.~V.~Nanopoulos,
  Phys.\ Lett.\ B {\bf 145}, 181 (1984).

    
  
  \bibitem{Juszkiewicz:gg}
R.~Juszkiewicz, J.~Silk and A.~Stebbins,
Phys.\ Lett.\ B {\bf 158} (1985) 463.

\bibitem{mmy}
T.~Moroi, H.~Murayama and M.~Yamaguchi,
  Phys.\ Lett.\ B {\bf 303}, 289 (1993).

 
  
\bibitem{Kawasaki:1994af} 
  M.~Kawasaki and T.~Moroi,
  Prog.\ Theor.\ Phys.\  {\bf 93}, 879 (1995)
  [hep-ph/9403364].
  
   \bibitem{Moroi:1995fs} 
  T.~Moroi,
  hep-ph/9503210.
  
  \bibitem{enor}
  J.~R.~Ellis, D.~V.~Nanopoulos, K.~A.~Olive and S.~J.~Rey,
  Astropart.\ Phys.\  {\bf 4}, 371 (1996)
  [hep-ph/9505438].
  
      \bibitem{Giudice:1999am} 
  G.~F.~Giudice, A.~Riotto and I.~Tkachev,
  JHEP {\bf 9911}, 036 (1999)
  [hep-ph/9911302].
  
   \bibitem{bbb}
   M.~Bolz, A.~Brandenburg and W.~Buchmuller,
  Nucl.\ Phys.\ B {\bf 606}, 518 (2001)
  [Erratum-ibid.\ B {\bf 790}, 336 (2008)]
  [hep-ph/0012052].
  
  \bibitem{cefo}
 R.~H.~Cyburt, J.~Ellis, B.~D.~Fields and K.~A.~Olive,
 Phys.\ Rev.\ D {\bf 67}, 103521 (2003) [astro-ph/0211258].


   \bibitem{kmy}
 K.~Kohri, T.~Moroi and A.~Yotsuyanagi,
  Phys.\ Rev.\ D {\bf 73}, 123511 (2006)
  [arXiv:hep-ph/0507245].

  
  
  \bibitem{stef}
    F.~D.~Steffen,
  JCAP {\bf 0609}, 001 (2006)
  [arXiv:hep-ph/0605306].
  
  \bibitem{Pradler:2006qh} 
  J.~Pradler and F.~D.~Steffen,
  Phys.\ Rev.\ D {\bf 75}, 023509 (2007)
  [hep-ph/0608344].
  
  \bibitem{ps2}
   J.~Pradler and F.~D.~Steffen,
  Phys.\ Lett.\ B {\bf 648}, 224 (2007)
  [hep-ph/0612291].
  
   \bibitem{kkmy}
  M.~Kawasaki, K.~Kohri, T~Moroi and A.Yotsuyanagi,
Phys.\ Rev.\ D {\bf 78}, 065011 (2008) [arXiv:0804.3745 [hep-ph]].
  
   \bibitem{rs}
   V.~S.~Rychkov and A.~Strumia,
  Phys.\ Rev.\ D {\bf 75}, 075011 (2007)
  [hep-ph/0701104].
 
  \bibitem{egnop}
  J.~Ellis, M.~A.~G.~Garcia, D.~V.~Nanopoulos, K.~A.~Olive and M.~Peloso,
  JCAP {\bf 1603}, no. 03, 008 (2016)
  [arXiv:1512.05701 [astro-ph.CO]].
  
  
   \bibitem{Garcia:2017tuj}
  M.~A.~G.~Garcia, Y.~Mambrini, K.~A.~Olive and M.~Peloso,
  Phys.\ Rev.\ D {\bf 96} (2017) no.10,  103510
  [arXiv:1709.01549 [hep-ph]].
  
   \bibitem{Giudice:2000ex}
  G.~F.~Giudice, E.~W.~Kolb and A.~Riotto,
  Phys.\ Rev.\ D {\bf 64} (2001) 023508
  [hep-ph/0005123];
   D.~J.~H.~Chung, E.~W.~Kolb and A.~Riotto,
  Phys.\ Rev.\ D {\bf 60} (1999) 063504
  [hep-ph/9809453].
  
 
  
  
  
    \bibitem{Kaneta:2019zgw}
  K.~Kaneta, Y.~Mambrini and K.~A.~Olive,
  Phys.\ Rev.\ D {\bf 99} (2019) no.6,  063508
  [arXiv:1901.04449 [hep-ph]].

  

\bibitem{Garcia:2020eof}
M.~A.~G.~Garcia, K.~Kaneta, Y.~Mambrini and K.~A.~Olive,
Phys. Rev. D \textbf{101}, no.12, 123507 (2020)
[arXiv:2004.08404 [hep-ph]].


\bibitem{Bernal:2020gzm}
N.~Bernal,
arXiv:2005.08988 [hep-ph].


\bibitem{ceflos}
M.~Kawasaki, F.~Takahashi and T.~T.~Yanagida,
Phys. Rev. D \textbf{74}, 043519 (2006)
[arXiv:hep-ph/0605297 [hep-ph]];
R.~H.~Cyburt, J.~Ellis, B.~D.~Fields, F.~Luo, K.~A.~Olive and V.~C.~Spanos,
  JCAP {\bf 0910}, 021 (2009)
  [arXiv:0907.5003 [astro-ph.CO]];
   R.~H.~Cyburt, J.~Ellis, B.~D.~Fields, F.~Luo, K.~A.~Olive and V.~C.~Spanos,
  JCAP {\bf 1305}, 014 (2013)
  [arXiv:1303.0574 [astro-ph.CO]].

  \bibitem{ego} 
  J.~L.~Evans, M.~A.~G.~Garcia and K.~A.~Olive,
  JCAP {\bf 1403}, 022 (2014)
  doi:10.1088/1475-7516/2014/03/022
  [arXiv:1311.0052 [hep-ph]].


\bibitem{Hidden}
See, for example,
B.~S.~Acharya, S.~A.~R.~Ellis, G.~L.~Kane, B.~D.~Nelson and M.~J.~Perry,
Phys. Rev. Lett. \textbf{117} (2016), 181802
[arXiv:1604.05320 [hep-ph]] and
JHEP \textbf{09} (2018), 130
[arXiv:1707.04530 [hep-ph]].

\bibitem{FY}
M.~Fukugita and T.~Yanagida,
Phys. Lett. B \textbf{174} (1986), 45-47.

\bibitem{AD}
I.~Affleck and M.~Dine,
Nucl. Phys. B \textbf{249} (1985), 361-380.

\bibitem{EKN2}
J.~R.~Ellis, C.~Kounnas and D.~V.~Nanopoulos,
Nucl. Phys. B \textbf{247}, 373-395 (1984)
doi:10.1016/0550-3213(84)90555-8

\end{thebibliography}
\end{document}